\pdfoutput=1

\documentclass[twocolumn,showpacs,amssymb,aps,nofootinbib,floatfix,superscriptaddress]{revtex4-1}

\usepackage{epsfig}
\usepackage{float}
\usepackage{amssymb}
\usepackage{hyperref}

\begin{document}

\title{Angle and magnitude decorrelation in the factorization breaking of collective flow}

\author{Piotr Bo\.zek}
\email{Piotr.Bozek@fis.agh.edu.pl}
\affiliation{AGH University of Science and Technology, Faculty of Physics and
Applied Computer Science, al. Mickiewicza 30, 30-059 Cracow, Poland}

\date{\today}

\begin{abstract}
The collective harmonic flow in heavy-ion collisions correlates particles
 at all transverse momenta to be emitted preferably some directions. 
The factorization breaking coefficient measures the small decorrelation 
of the flow harmonics at two different transverse momenta. Using
 the hydrodynamic model I study 
in details the decorrelation of the harmonic flow due to the flow angle 
and  the flow magnitude decorrelation at two  transverse momenta.
The effect can be seen in experiment measuring
 factorization breaking coefficients
for the square of the harmonic flow vector at two  transverse momenta.
The hydrodynamic model predicts that the decorrelation of the  flow
magnitudes is about one half of the decorrelation of the overall flow  
(combining
 flow angle and flow magnitude decorrelations). These results are consistent
 with the principal component analysis of correlators of flow 
vectors squared.
\end{abstract}




\maketitle

\section{Introduction}

The collective expansion of dense matter created in relativistic nuclear
collisions creates strong correlation between emitted particles 
\cite{Heinz:2013th,Gale:2013da,Ollitrault:2010tn}.
The azimuthal asymmetry of the collective flow gives rise to an asymmetry of
particle spectra
\begin{eqnarray*}
\frac{dN}{d\phi} &\propto & 1+ v_2 \cos\left( 2(\phi -\Psi_2)\right)
\nonumber \\
& + & v_3 \cos\left( 3(\phi -\Psi_3)\right) + \dots \ .
\end{eqnarray*}
In the above equation  the elliptic $v_2$
 and triangular $v_3$ flows and the flow angles $\Psi_n$ are collective parameters of the spectra that fluctuate from event to event. The harmonic flow coefficients can be extracted from two (or higher) particle correlations.
The study of the flow coefficients in heavy-ion collisions at different 
centralities is a way to extract the properties of the expanding medium,
 in particular the value of shear viscosity.

The collective parameters can depend  on particle transverse momentum 
or pseudorapidity
$v_n(p_\perp,\eta)$, $\Psi_n(p_\perp,\eta)$. Using correlations of 
two particles at different pseudorapidities \cite{Bozek:2010vz} or different
transverse momenta \cite{Gardim:2012im}
the decorrelation of the flow parameters at different $p_\perp$ or $\eta$ can be observed. 
The phenomenon is known as flow  factorization breaking in pseudorapidity 
or transverse momentum. 
 It has been measured  experimentally 
\cite{Acharya:2017ino,Aad:2014lta,Khachatryan:2015oea,Aaboud:2017tql} 
and calculated in models 
\cite{Gardim:2012im,Heinz:2013bua,Kozlov:2014fqa,Lin:2004en,Pang:2015zrq,Xiao:2015dma}. It is found that the factorization breaking coefficient 
is sensitive to  fluctuations in the initial state. The initial
fluctuations are transformed by the collective expansion into a small 
decorrelation of  flow parameters at two transverse momenta 
or pseudorapidities. The predicted decorrelation is not strongly
 dependent on the viscosity.

Factorization breaking coefficients from two-particle correlations
measure the overall flow decorrelation, which is a
 combined effect of the decorrelation of the collective flow
 magnitudes $v_n(p_\perp,\eta)$ and of the decorrelation of the flow angles
$\Psi_n(p_\perp,\eta)$ at two different transverses momenta or rapidities.
By using $4$-particle correlators the flow angle decorrelation and 
the overall flow decorrelation in pseudorapidity 
can be measured separately \cite{Aaboud:2017tql,Jia:2017kdq}.
Experimental 
data and hydrodynamic model simulations \cite{Bozek:2017qir,Wu:2018cpc} show
that the angle decorrelation accounts for only a part of the overall
 flow decorrelation in pseudorapidity.

In this paper the analogous effect is studied for the harmonic
 flow decorrelation in transverse momentum. In the hydrodynamic model the 
flow angle and the flow magnitude decorrelations 
 at different transverse momenta are studied separately and compared 
to the overall flow decorrelation (Sect. \ref{sec:angmag}). 
Model calculations show that the flow decorrelation in a particular 
event is correlated to the flow magnitude in the same event. 
This could be measured experimentally by using  correlators
weighed with different powers of $v_n$
 (Sect. \ref{sec:decmag}). In order to measure the flow angle 
or flow magnitude decorrelations separately $4$-particle 
correlations must be used. Unlike for the decorrelation in pseudorapidity,
 where the flow angle decorrelation can be estimated, for the factorization breaking
 in 
transverses momentum a measure of the flow magnitude decorrelation 
can be naturally defined (Sect. \ref{sec:flowmagn}). The 
flow magnitude decorrelation accounts for about one half of the overall flow decorrelation at two different transverse momenta.
The findings are consistent with the principal component analysis of 
 correlation matrices of higher powers of harmonic flow (Sect. \ref{sec:pca}).

\section{Model \label{sec:model}}

I  use  3+1 dimensional relativistic 
viscous hydrodynamics  with Monte Carlo
 Glauber model initial conditions  ~\cite{Schenke:2010rr,Bozek:2011ua} to model Pb+Pb collisions at $\sqrt{s_{NN}}=5.02$~TeV.  
 The initial entropy deposition 
in the transverse plane is given as a sum of contributions from  
$N_p$  participant nucleons
\begin{equation}
s(x,y)=\sum_{i=1}^{N_p} g_i(x,y)   \  \label{eq:entropy}
\end{equation} 
(the form of the initial distribution in the longitudinal direction is 
skipped here for simplicity, it follows the parametrization given 
in Ref. \cite{Bozek:2011ua}).
Each nucleon gives a Gaussian-smeared  contribution
\begin{eqnarray}
g_i(x,y)& =  & \kappa \left( (1-\alpha)+\alpha c_i \right) \nonumber \\
& &  \exp\left( - \frac{(x-x_i)^2+(y-y_i)^2}{2\sigma^2}\right) , 
\end{eqnarray} 
where $c_i$ is the number of collisions for nucleon $i$ at position
$(x_i,y_i)$ and $\kappa$ is adjusted to reproduce the charged particle density in pseudorapidity. For Pb+Pb collisions at $\sqrt{s_{NN}}=5.02$~TeV $\alpha=0.1$ is taken to describe the centrality dependence of the charged particle density.
At the freeze-out temperature $150$~MeV  statistical emission of 
hadrons takes place \cite{Chojnacki:2011hb}.

For each hydrodynamic event I generate $150$ or $600$  events for  centralities $0-5$\% and $30-40$\%.  For each hydrodynamic event the collective flow is calculated by combining these events. This procedure
 reduces statistical errors and non-flow contributions 
in multiparticle correlators \cite{Bozek:2017thv}. 
The event-by-event reconstruction of the flow vectors in the model 
allows to  calculate the flow angle or flow magnitude decorrelation 
separately. The flow correlations are analyzed in the transverse momentum range 
$p_\perp \in [0.3,3.0]$~GeV. The $p_\perp$ range is divided into  bins of unequal width, but of equal mean particle multiplicity in each bin
 \cite{Bozek:2017thv}. This choice guarantees that the statistical
 errors for the  all the elements of  correlators in two $p_\perp$ bins are similar.

\section{Flow angle and flow magnitude decorrelation \label{sec:angmag}}

\begin{figure}
\begin{center}
\includegraphics[angle=0,width=0.41 \textwidth]{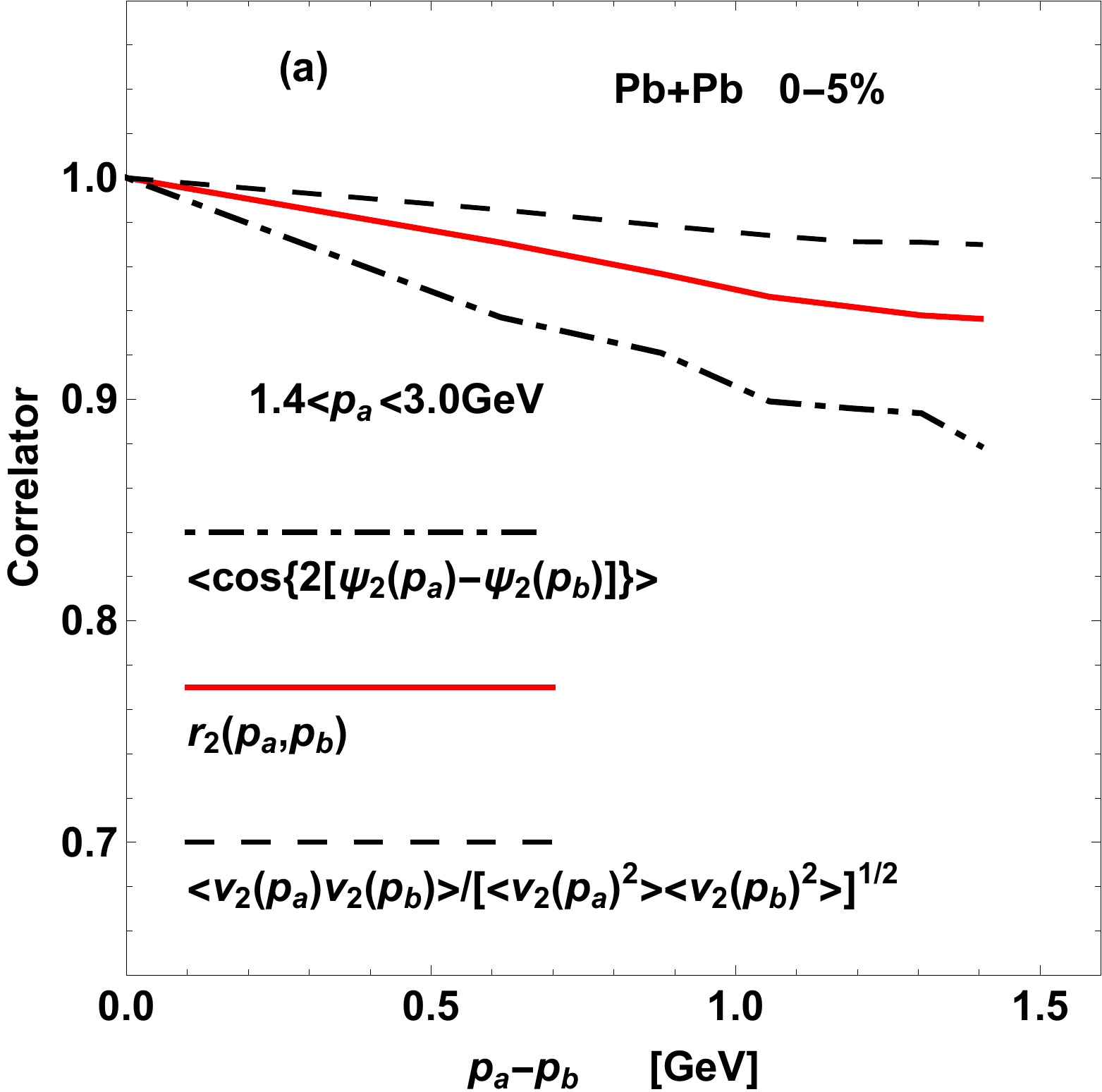}
\vspace{4mm}  

\includegraphics[angle=0,width=0.41 \textwidth]{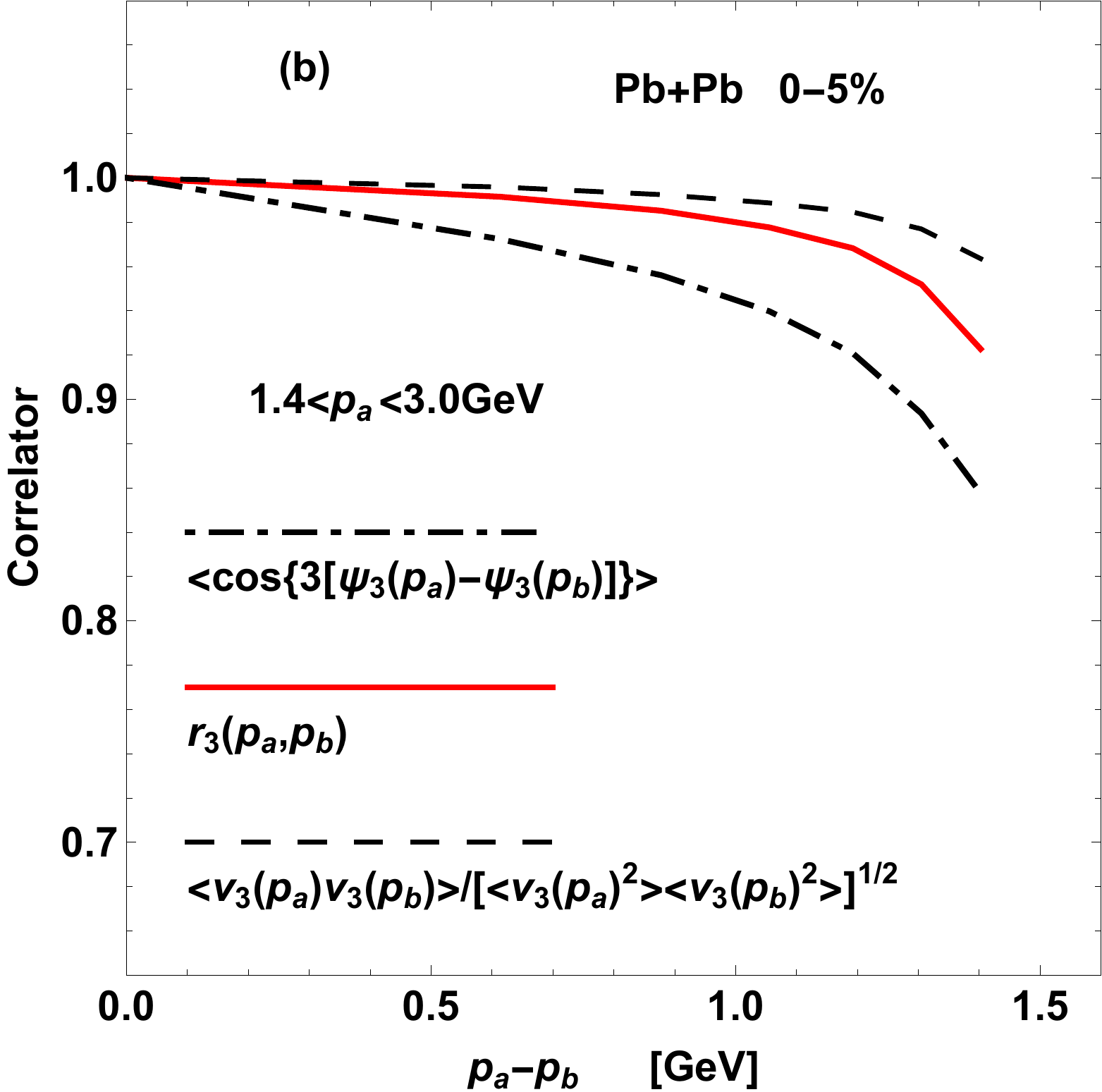}  
\end{center}
\vspace{-5mm}
\caption{Correlator for the flow angle from Eq.~(\ref{eq:dvnang}) (dot-dashed line), for the flow magnitude from Eq.~(\ref{eq:dvnmag}) 
(dashed line), and for the harmonic flow  from Eq.~(\ref{eq:rn})  (solid line)
 for two bins in transverse momentum. 
 Panels (a) and (b) show the results for the second- and third-order harmonic
flow, respectively. Pb+Pb collisions at $\sqrt{s_{NN}}=5.02$~TeV for centrality 0-5\%.
\label{fig:vam05}} 
\end{figure}

\begin{figure}
\begin{center}
\includegraphics[angle=0,width=0.41 \textwidth]{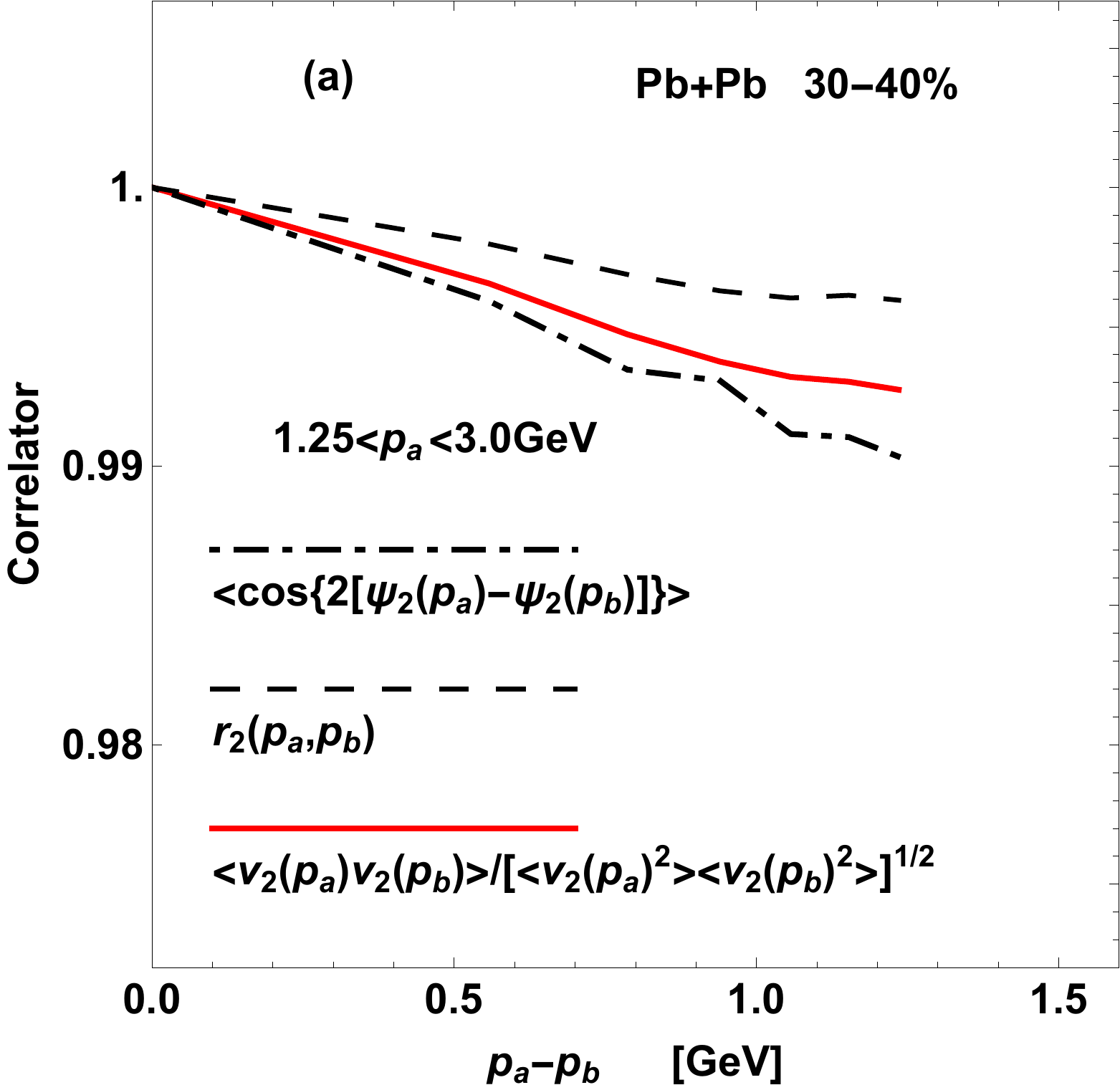}  
\vspace{4mm}

\includegraphics[angle=0,width=0.41 \textwidth]{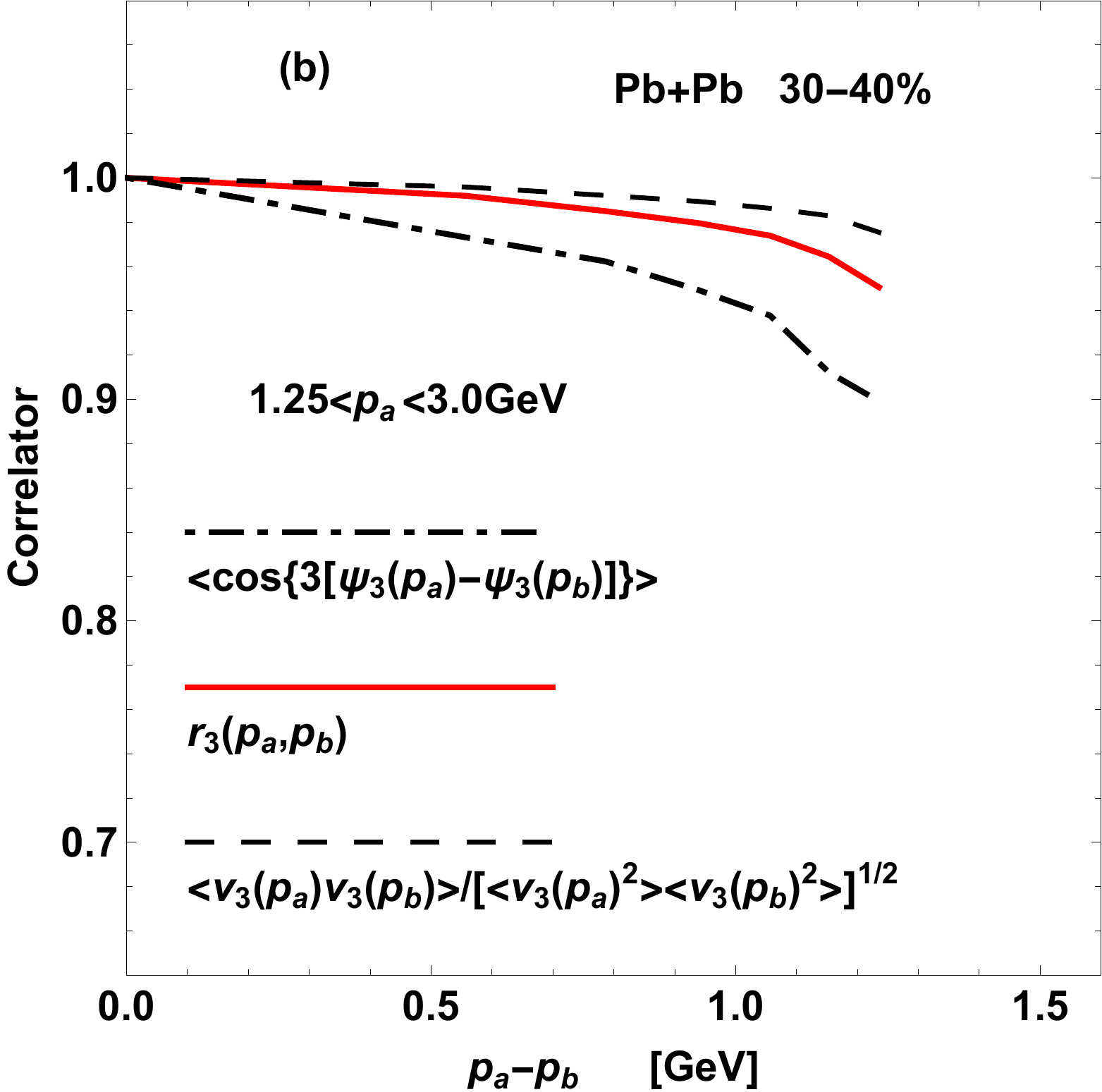}  
\end{center}
\vspace{-5mm}
\caption{Same as Fig. \ref{fig:vam05} but for centrality 30-40\%.
\label{fig:vam3040}} 
\end{figure}

The decorrelation  of   harmonic flow at two different
 transverse momenta $p_a$ and $p_b$ 
is measured using the factorization breaking 
coefficient \cite{Gardim:2012im}
\begin{equation}
r_n(p_a,p_b)=\frac{\langle q_n(p_a) q_n^\star(p_b) \rangle}
{\sqrt{\langle v_n(p_a)^2 \rangle \langle v_n(p_p)^2 \rangle }}  \ ,
\label{eq:rn}
\end{equation}
where  
\begin{equation}
q_n(p) = \frac{1}{N} \sum_{j} e^{ i n \phi_j} = v_n(p) e^{i n \Psi_n(p)} 
\end{equation}
is the $q$ vector of the n-th order harmonic flow calculated from the azimuthal
 angles $\phi_j$ of $N$ particles in the bin at transverse momentum $p$
and $\langle \dots \rangle$ denotes the average over events.
The n-th harmonic at transverse momentum $p$ can be written
as 
\begin{equation}
\langle v_n(p)^2 \rangle = \langle q_n(p) q_n^\star(p) \rangle 
=\frac{1}{N(N-1)}\sum_{j\neq k}  e^{i n (\phi_j-i  \phi_k)} \ .
\end{equation}
In this paper I use  the convention that selfcorrelation terms 
are dropped from 
sums over particles in  the same bin.

For flow dominated correlations between emitted particles 
the factorization breaking coefficient $r_n$ measures the correlation coefficient of the
flow vectors $q_n$ at different transverse momenta. In that case $r_n(p_a,p_b)\le 1$ \cite{Gardim:2012im}. The value $r_n(p_a,p_b) <1$ means that the harmonic 
flow at the 
transverse momenta $p_a$ and $p_b$ is partially decorrelated. 
This decorrelation can be due to a flow magnitude or a flow angle 
decorrelation \cite{Jia:2014vja}. Flow angle decorrelation means that event-by-event differences in the effective flow angles $\Psi_n(p_a)$ and $\Psi_n(p_b)$ at the two transverse momenta appear. The flow angle difference 
$\Psi(p_a)-\Psi_n(p_b)$ contributes a factor 
$\cos\left\{  n \left[\Psi_n(p_a) -\Psi_n(p_b)\right] \right\}$ in the numerator of of the factorization breaking coefficient $r_n(p_a,p_b)$.
The decorrelation of the  harmonic flow angles
is defined as
\begin{equation}
\langle \cos\left \{  n \left[\Psi_n(p_a) -\Psi_n(p_b)\right] \right \} \rangle \ .
\label{eq:dvnang} 
\end{equation}

 The decorrelation of the magnitude of the harmonic flow at two transverse momenta
can be defined as
\begin{equation}
\frac{\langle v_n(p_a) v_n(p_b)\rangle}{\sqrt{\langle v_n(p_a)^2 
\rangle \langle v_n(p_p)^2 \rangle }}  \ .
\label{eq:dvnmag}
\end{equation}
Please note that 
the angle (\ref{eq:dvnang}) and magnitude (\ref{eq:dvnmag}) decorrelations 
 cannot be calculated from experimental data.
On the other hand, these quantities can be estimated in the hydrodynamic model 
integrating over the particle distributions in momenta, instead of a summation 
over particles in an event. In practice this integration is performed using a Monte Carlo method by generating a large number of particles at the freeze-out hypersurface, as described in Sect. \ref{sec:model}.
If the angle and magnitude decorrelation factorize, on has
\begin{eqnarray}
& & r_n(p_a,p_b)\simeq   \nonumber \\
& & \frac{\langle v_n(p_a) v_n(p_b)\rangle}{\sqrt{\langle v_n(p_a)^2 
\rangle \langle v_n(p_p)^2 \rangle }}\langle \cos\left \{  n \left[\Psi_n(p_a) -\Psi_n(p_b)\right] \right \} \rangle \ .
\label{eq:appfac}
\end{eqnarray}

In Figs. \ref{fig:vam05} and \ref{fig:vam3040} are compared the factorization breaking coefficients $r_n(p_a,p_b)$, the angle decorrelation (\ref{eq:dvnang}), and the magnitude decorrelation (\ref{eq:dvnmag}). The factorization breaking 
coefficient is not a simple product of the angle and magnitude decorrelations as in Eq. \ref{eq:appfac}. In fact, an inverted hierarchy of decorrelations appears.
The angle decorrelation (\ref{eq:dvnang}) is stronger than the flow decorrelation
 given by the factorization breaking coefficient $r_2$.
The reason for the inverted hierarchy is that the three averages 
in Eq. \ref{eq:appfac} are weighted with different powers of $v_n$ \cite{Bozek:2017qir}.  
The decorrelation of the flow angles is anticorrelated with the overall
 magnitude of the flow in an event. Therefore the average (\ref{eq:dvnang}),
that is weighted with a zeroth power of $v_n$, gives a larger deviation from $1$, i.e.
 a stronger decorrelation than the other 
two averages in Figs. \ref{fig:vam05} and
 \ref{fig:vam3040}.

\section{Correlation of the overall flow magnitude and of the flow decorrelation\label{sec:decmag}}

On an event-by-event basis an anti-correlation occurs between the magnitude of the flow $v_n^2$ in the event and  the factorization breaking coefficient.
In events with a larger flow the the decorrelation 
is smaller ($r_n$ is bigger,
closer to $1$). This effect has been observed in model calculations for the decorrelation in pseudorapidity \cite{Bozek:2017qir}. The same effect
 can be evidenced for the decorrelation of flow in transverse momentum.

Experimentally  factorization breaking coefficients weighted with different
powers of $v_n$ can be defined
\begin{equation}
r_{n|n;1}^{n;2k}(p_a,p_b)=\frac{\langle v_n^{2k} q_n(p_a) q_n^\star(p_b) \rangle}
{\sqrt{\langle v_n^{2k} v_n(p_a)^2 \rangle \langle v_n^{2k} v_n(p_p)^2 \rangle }} \ .
\label{eq:rvk}
\end{equation}
For $k=0$ the above formula reduces to the standard factorization breaking coefficient $r_n(p_a,p_b)$.
The correlators in the numerator and denominator of  Eq. \ref{eq:rvk} involve 
summation over $2+2k$ particles. Self-correlations must be subtracted in the summation. In experiment non-flow effects can be reduced using rapidity gaps.

\begin{figure}
\begin{center}
\includegraphics[angle=0,width=0.41 \textwidth]{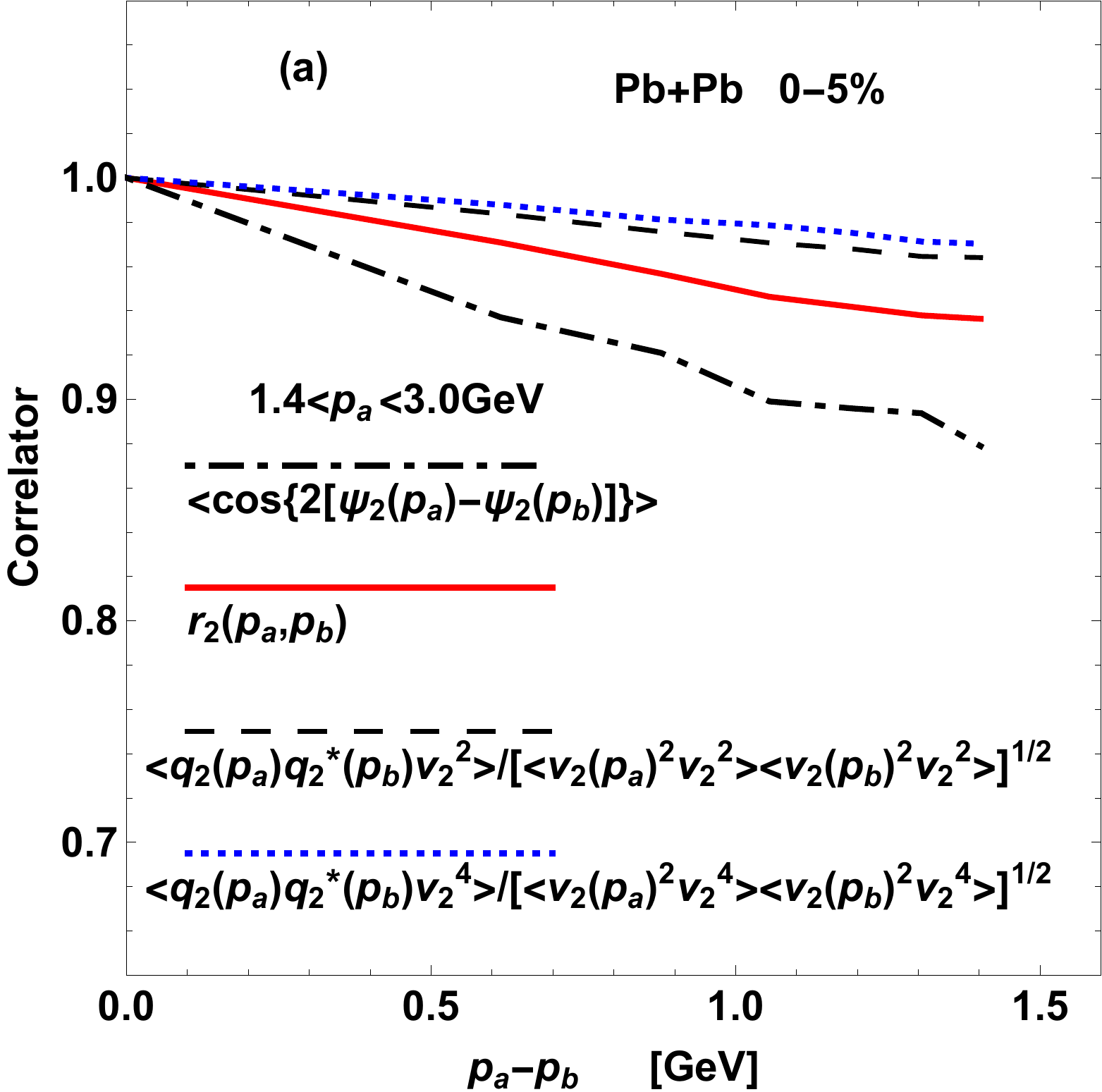}
\vspace{4mm}  

\includegraphics[angle=0,width=0.41 \textwidth]{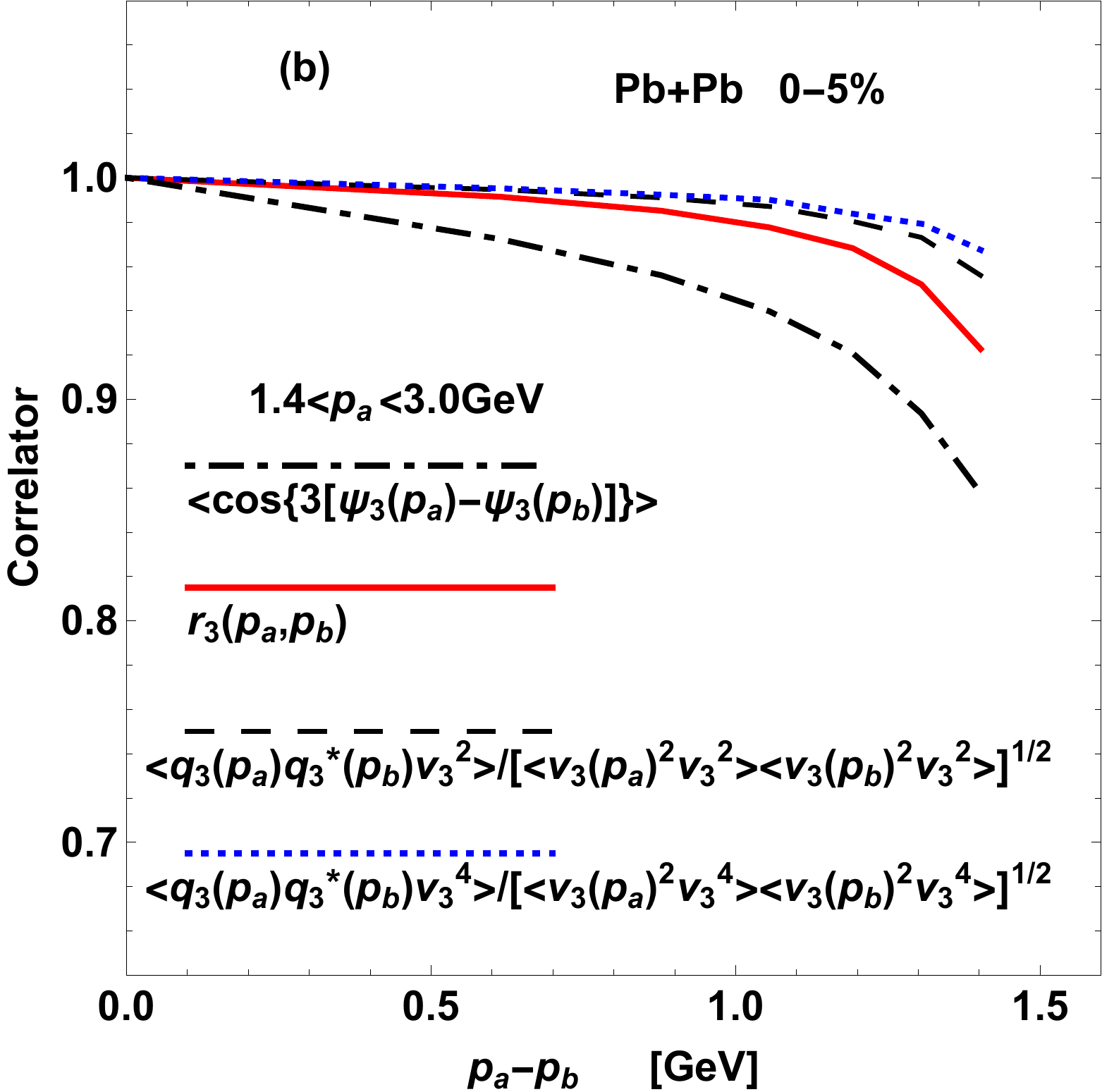}  
\end{center}
\vspace{-5mm}
\caption{Correlator of the flow  (solid line), of the flow weighted with $v_n^2$  (dashed line), and of the flow weighted with $v_n^4$ (dotted line) at two different transverse momenta. The dash-dotted line represent the correlator for the flow angle (\ref{eq:dvnang}).  Panels (a) and (b) show the results for the second- and third-order harmonic
flow, respectively. Pb+Pb collisions at $\sqrt{s_{NN}}=5.02$~TeV for centrality 0-5\%.
\label{fig:vv05}}
\end{figure}

\begin{figure}
\begin{center}
\includegraphics[angle=0,width=0.41 \textwidth]{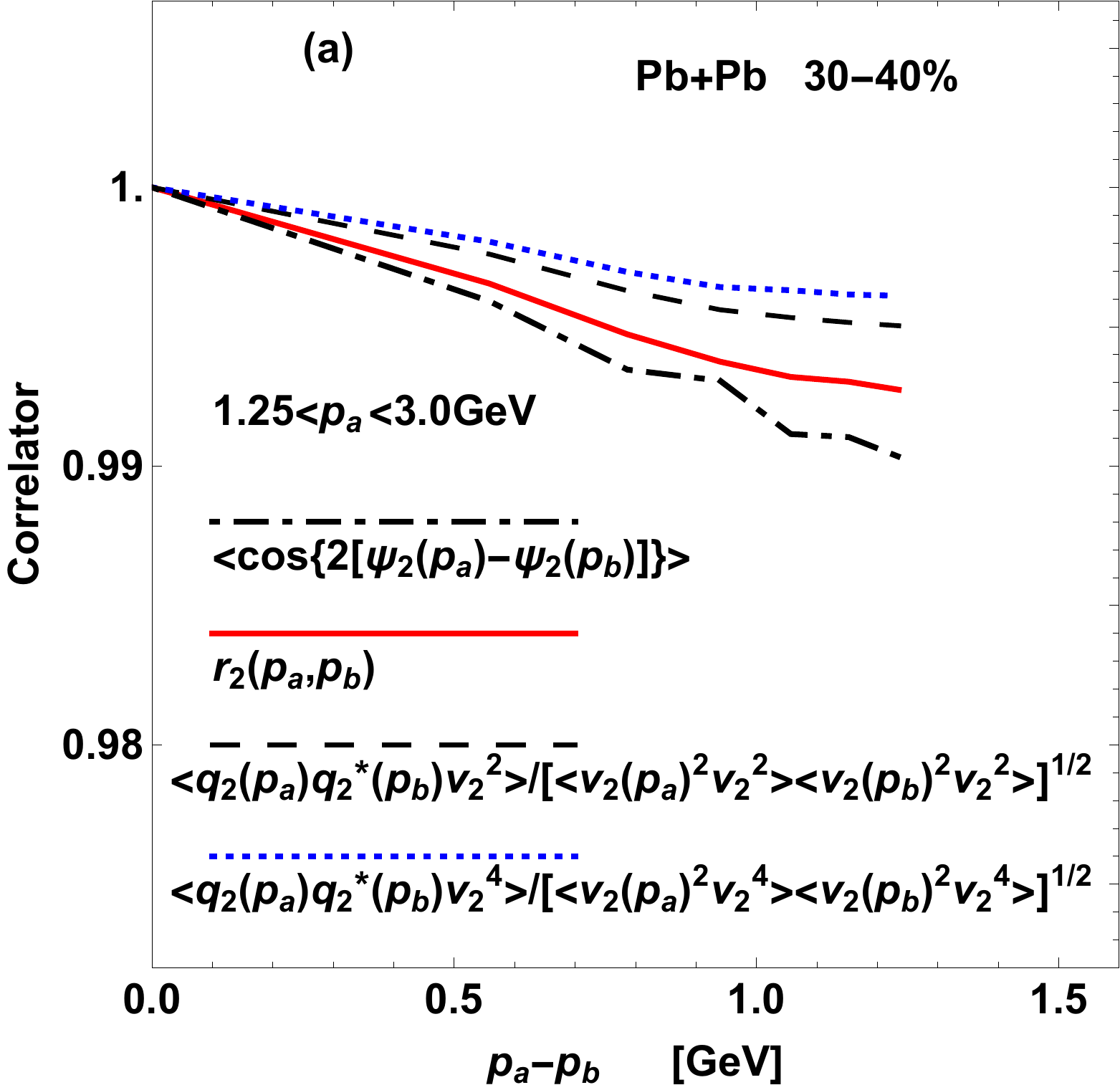}  
\vspace{4mm}

\includegraphics[angle=0,width=0.41 \textwidth]{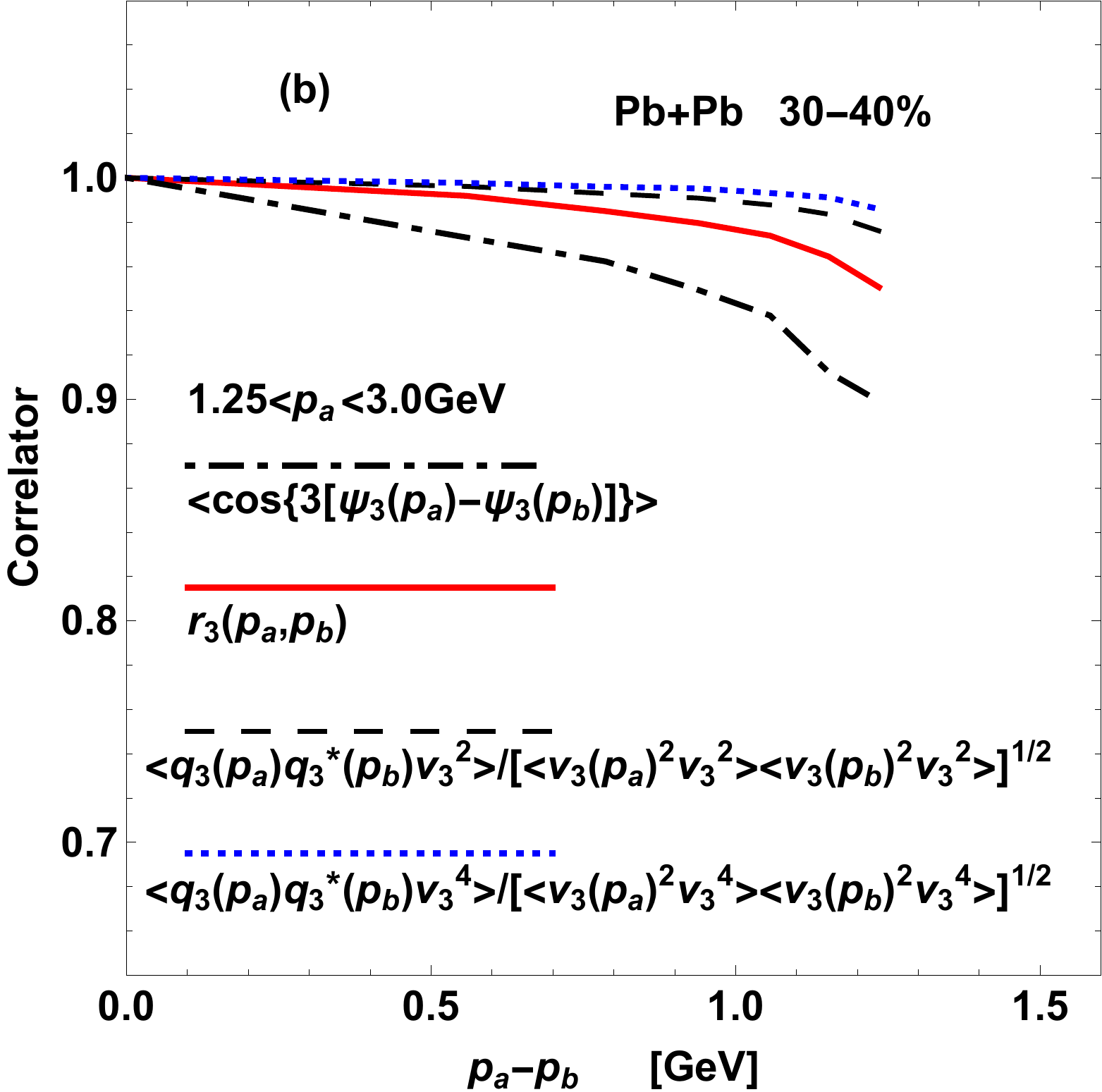}  
\end{center}
\vspace{-5mm}
\caption{Same as Fig. \ref{fig:vv05} but for centrality 30-40\%.
\label{fig:vv3040}} 
\end{figure}

In Figs. \ref{fig:vv05} and \ref{fig:vv3040} are shown the correlators
$r_{n|n;1}^{n;2k}(p_a,p_b)$ for $k=0,1,2$. The flow magnitude in the weighting factor $v_n^{2k}$ corresponds to the integrated flow in $p_\perp \in [0.3,3.0]$~Gev.
For correlators with higher powers of the weighting factor $v_n^{2k}$ the decorrelation is weaker.  This prediction could be
tested in experiment in order to evidence the relation between the overall 
flow magnitude and the flow decorrelation in transverse momentum.
The angle decorrelation (\ref{eq:dvnang}) is shown in Figs.  \ref{fig:vv05} and \ref{fig:vv3040} as well (dash-dotted lines). This quantity give the strongest decorrelation, as it corresponds to having effectively a weighting factor $v_n^{-2}$. This last correlator can be estimated in the model  but not in the experiment.

\section{Higher order flow correlators}

\begin{figure}
\begin{center}
\includegraphics[angle=0,width=0.41 \textwidth]{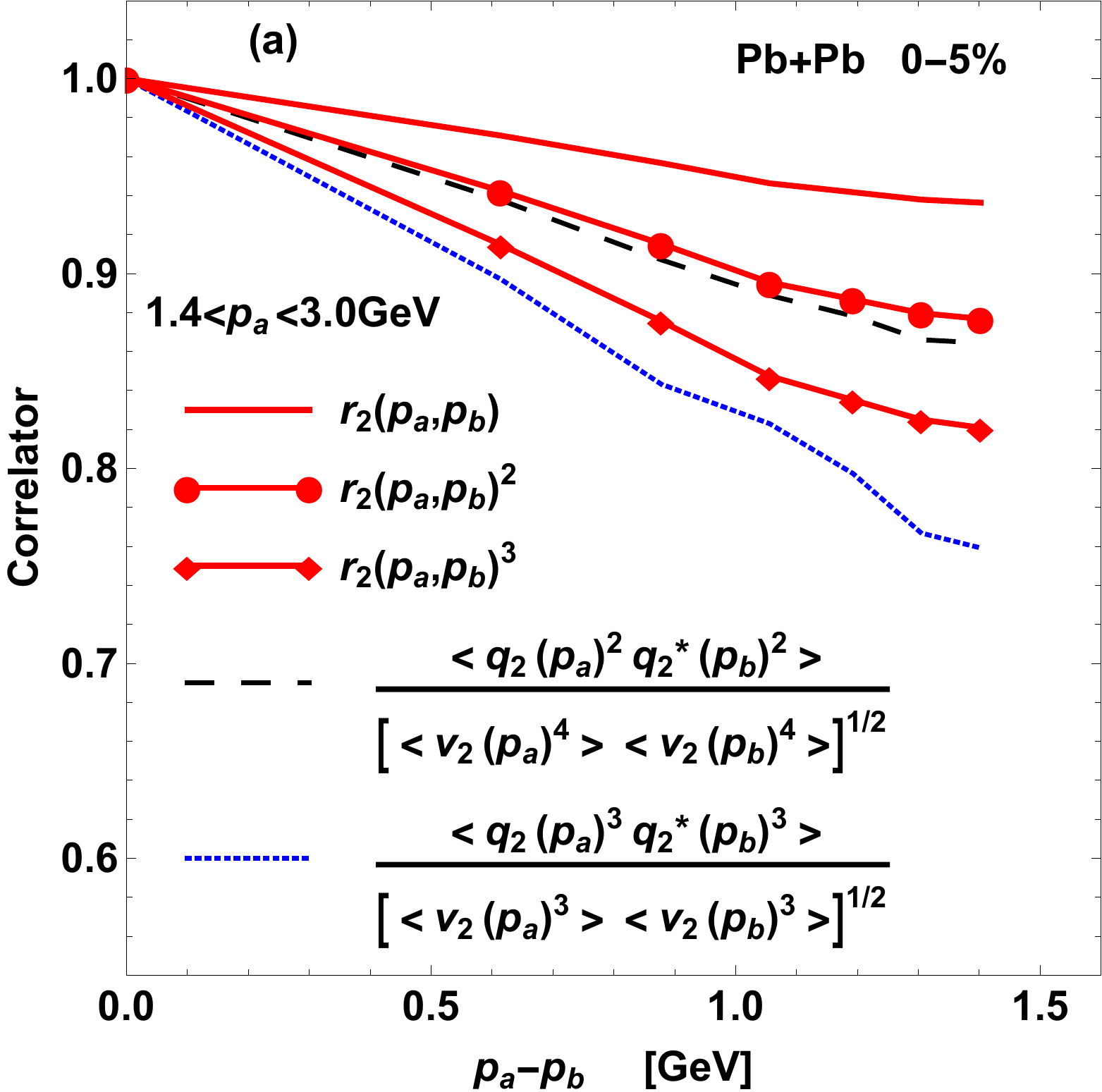}
\vspace{4mm}  

\includegraphics[angle=0,width=0.41 \textwidth]{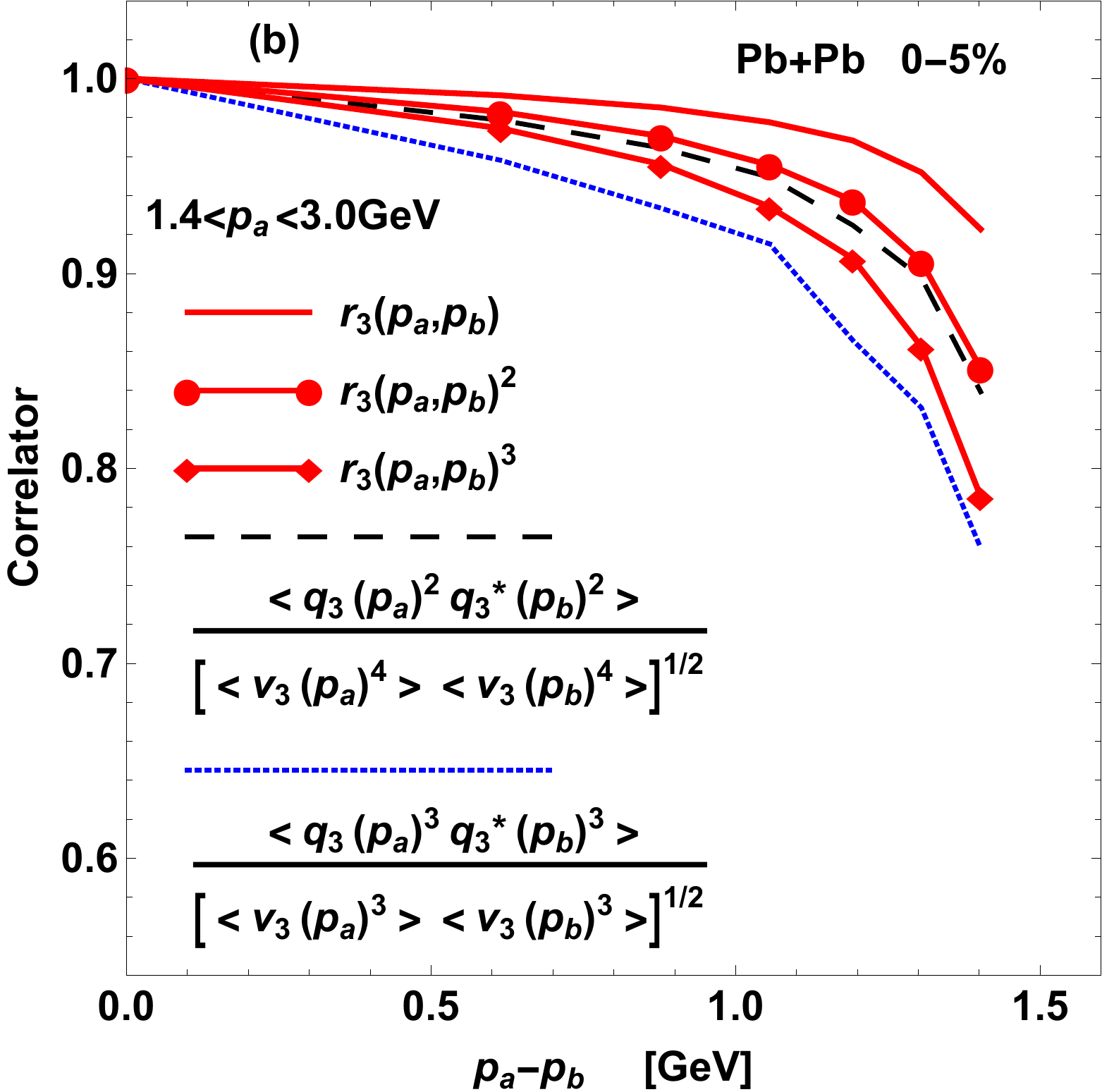}  
\end{center}
\vspace{-5mm}
\caption{Correlator of the flow  (solid line), of the flow squared (dashed line), and of the flow to power $3$ (dotted line) at two different transverse momenta. Solid lines with symbols represent second and third powers of the standard flow correlator $r_2$.  Panels (a) and (b) show the results for the second- and third-order harmonic
flow, respectively.  Pb+Pb collisions at $\sqrt{s_{NN}}=5.02$~TeV for centrality 0-5\%.
\label{fig:qq05}} 
\end{figure}

\begin{figure}
\begin{center}
\includegraphics[angle=0,width=0.41 \textwidth]{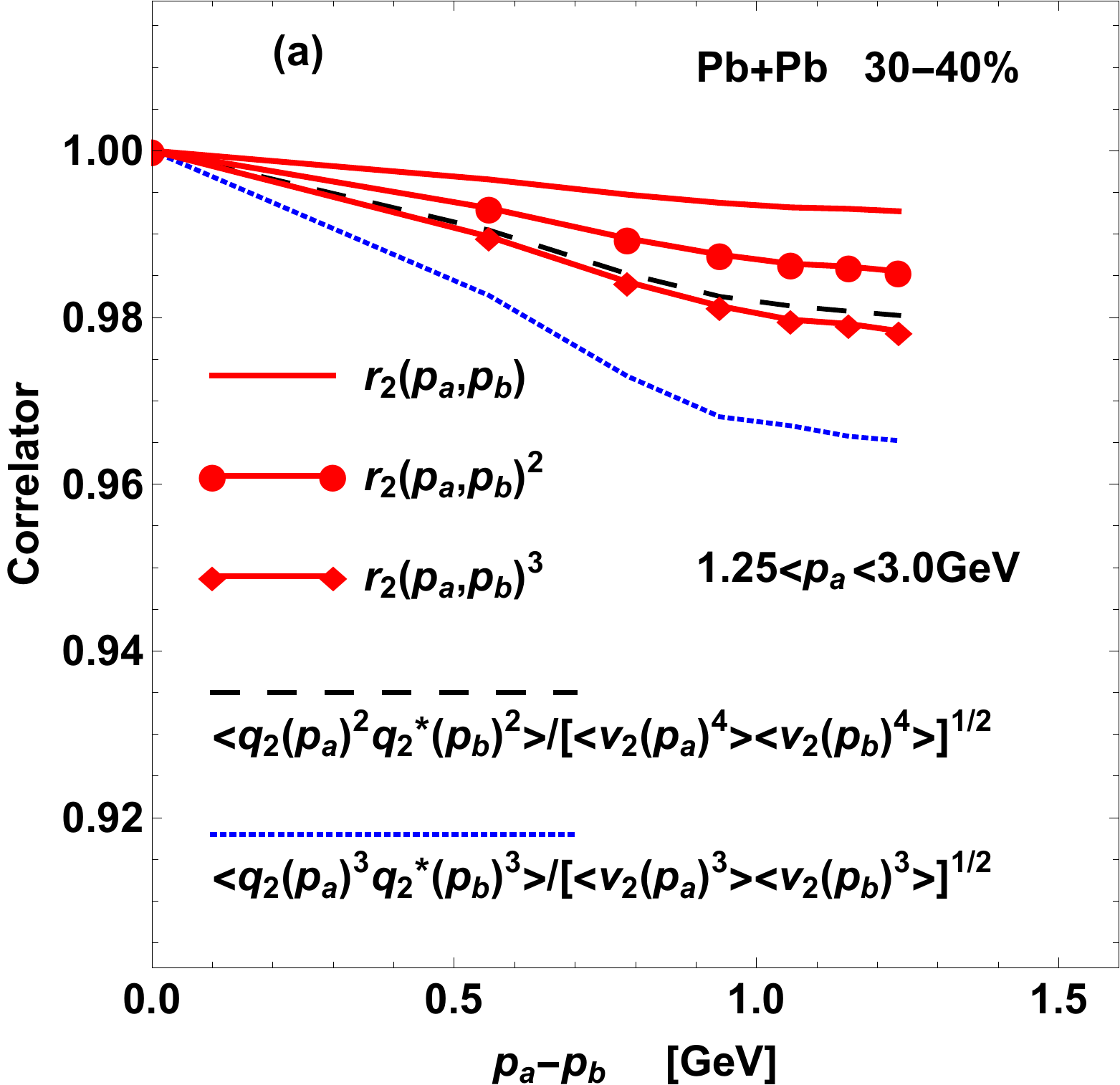}  
\vspace{4mm}

\includegraphics[angle=0,width=0.41 \textwidth]{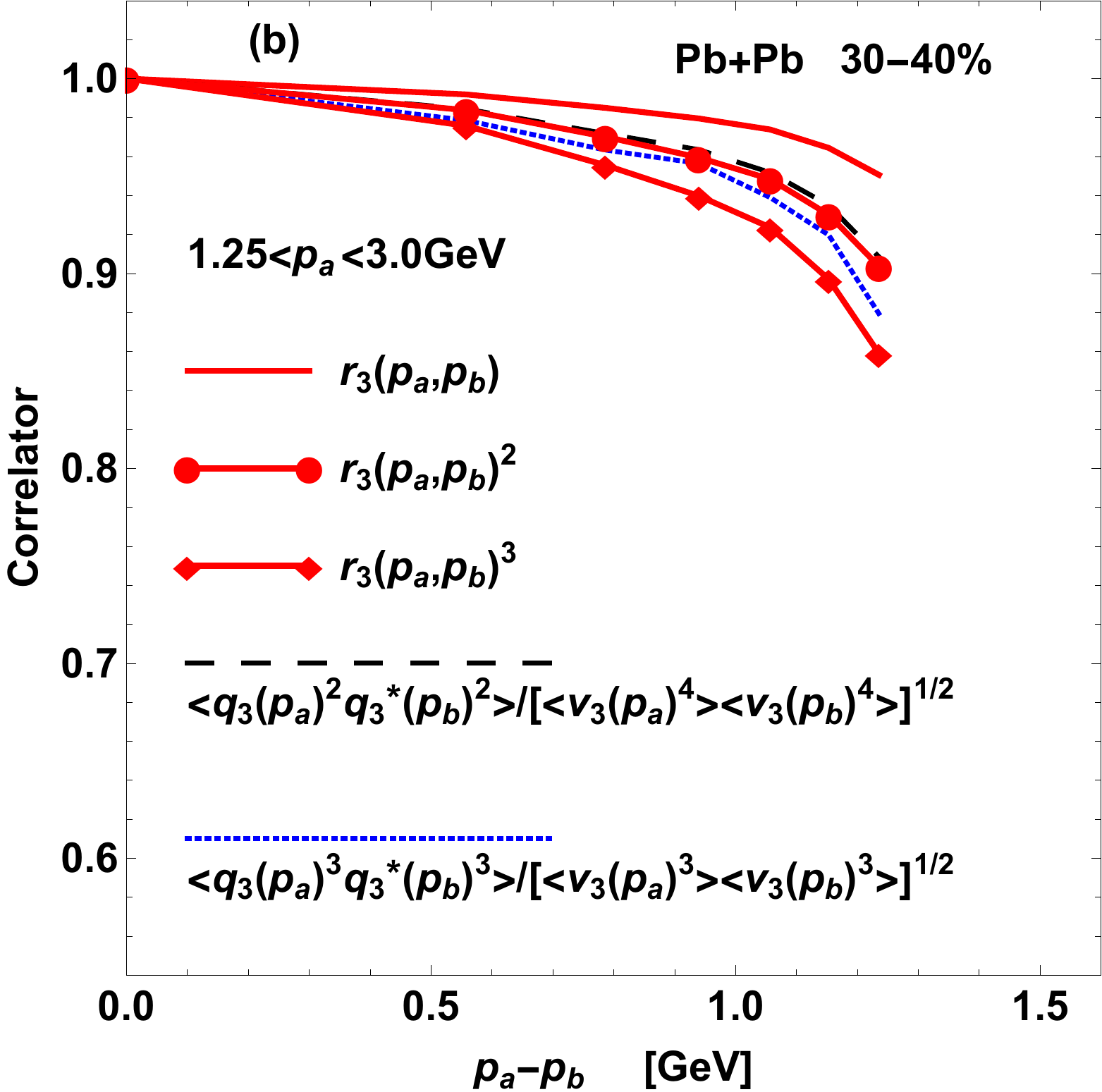}  
\end{center}
\vspace{-5mm}
\caption{Same as Fig. \ref{fig:qq05} but for centrality 30-40\%.
\label{fig:qq3040}} 
\end{figure}

Correlators of higher powers of the flow in two
 different pseudorapidity bins have 
been measured experimentally \cite{Aaboud:2017tql} and calculated in the hydrodynamic model \cite{Bozek:2017qir,Wu:2018cpc}. The simplest higher order correlators involve higher powers of the $q$ vectors
\begin{equation}
r_{n|n;k}(p_a,p_b)=\frac{\langle q_n(p_a)^k q_n^\star(p_b)^k \rangle}
{\sqrt{\langle v_n(p_a)^{2k} \rangle \langle v_n(p_p)^{2k} \rangle }} \ .
\label{eq:hiq}
\end{equation}
For $k=1$ one recovers the factorization breaking coefficient (\ref{eq:rn})
$r_{n|n;1}(p_a,p_b)=r_n(p_a,p_b)$. 

Correlators involving higher powers of the flow vector yield stronger decorrelation,
$r_{n|n;3}(p_a,p_b)<r_{n|n;2}(p_a,p_b)<r_{n|n;1}(p_a,p_b)$ (Figs. \ref{fig:qq05} and \ref{fig:qq3040}).
Factorization breaking coefficients for higher  powers of flow vectors do not factorize
into powers of the basic factorization coefficient 
\begin{equation}
r_{n|n;k}(p_a,p_b)\neq r_{n}(p_a,p_b)^k \ \ . 
\end{equation}

\section{Measuring flow magnitude factorization breaking \label{sec:flowmagn}}

\begin{figure}
\begin{center}
\includegraphics[angle=0,width=0.41 \textwidth]{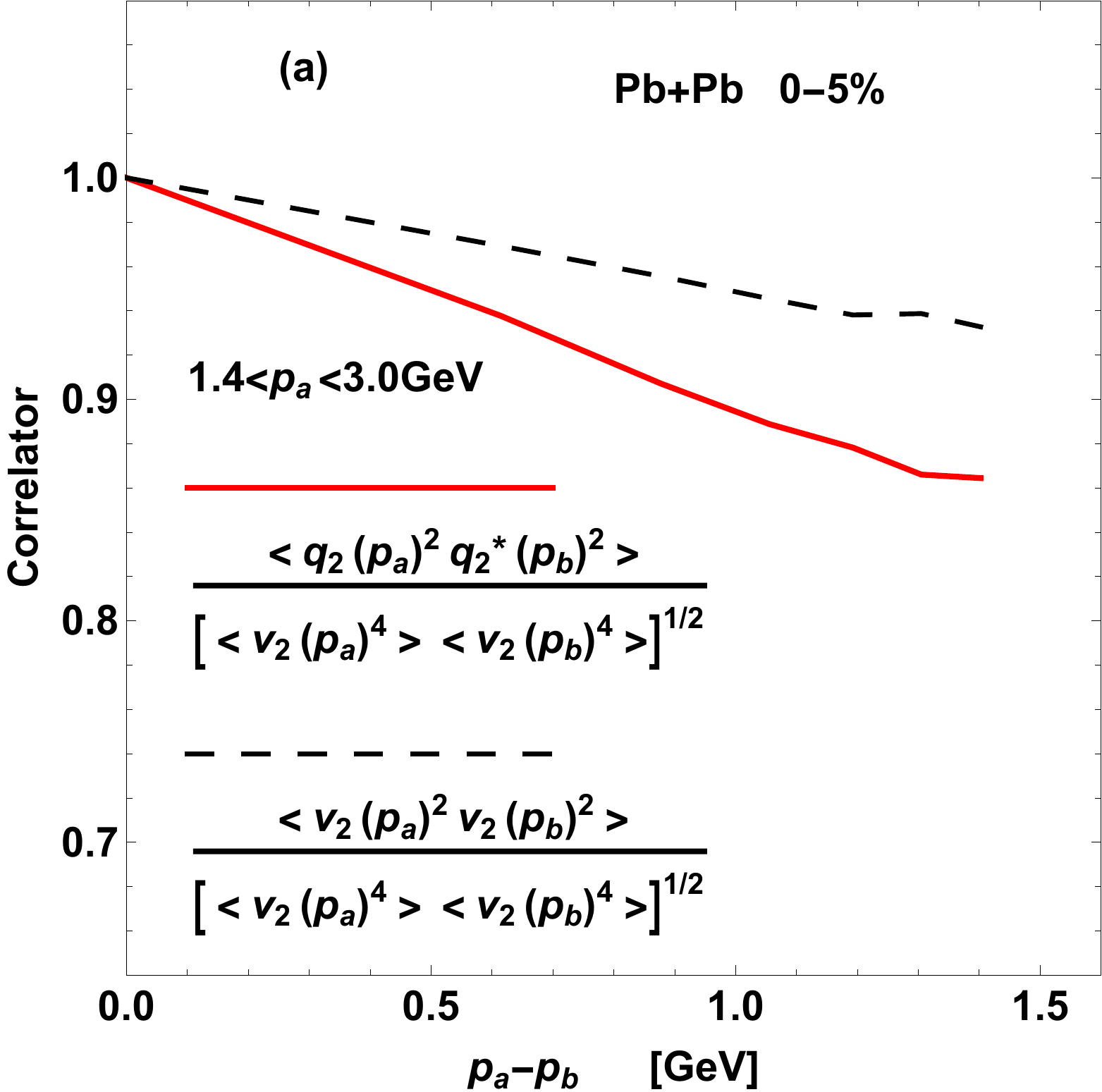}
\vspace{4mm}  

\includegraphics[angle=0,width=0.41 \textwidth]{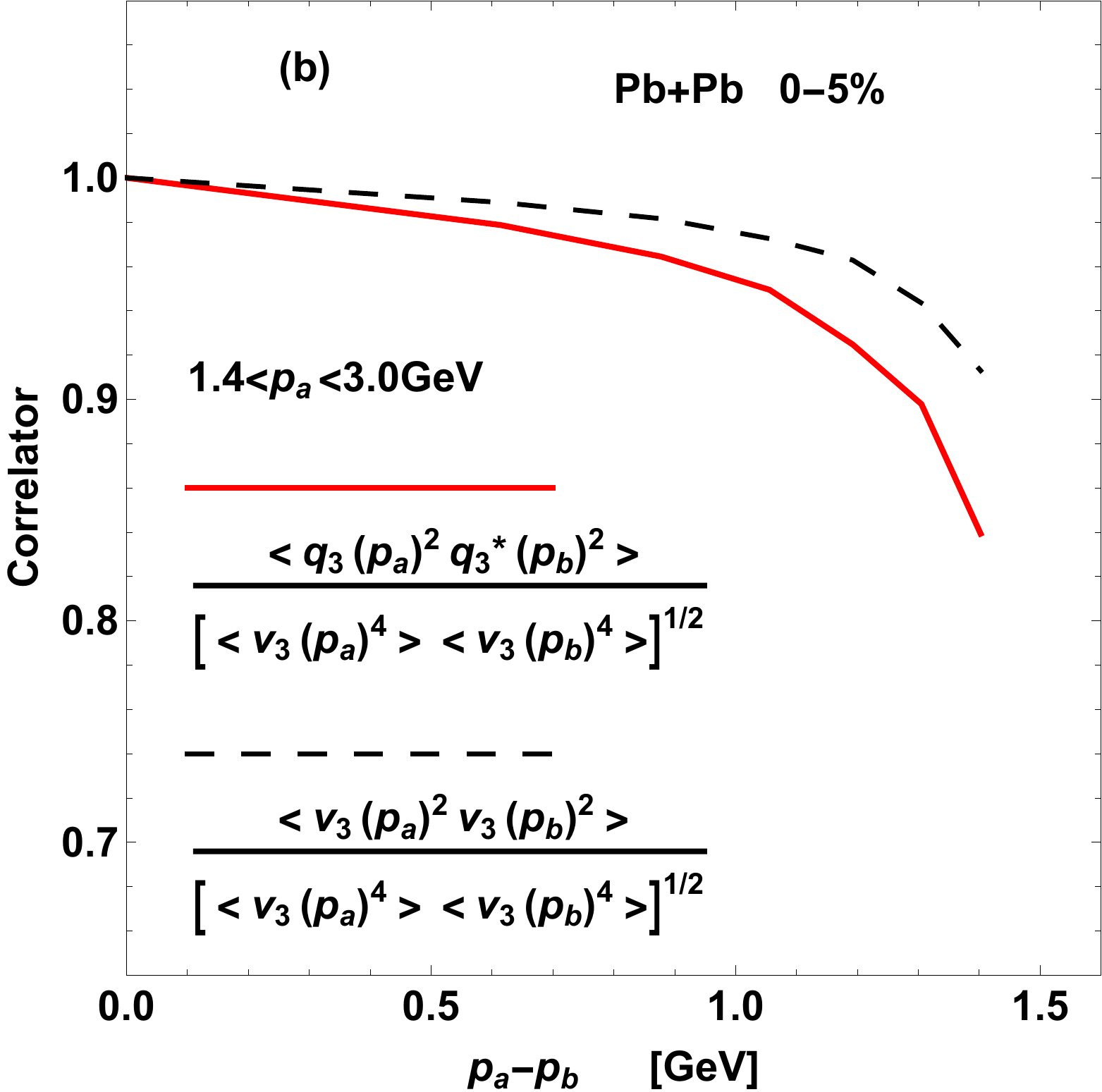}  
\end{center}
\vspace{-5mm}
\caption{Correlator for the flow squared  (solid line) and for the flow magnitude squared 
(dashed line).  Panels (a) and (b) show the results for the second- and third-order harmonic
flow, respectively. Pb+Pb collisions at $\sqrt{s_{NN}}=5.02$~TeV for centrality 0-5\%.
\label{fig:k205}} 
\end{figure}

\begin{figure}
\begin{center}
\includegraphics[angle=0,width=0.41 \textwidth]{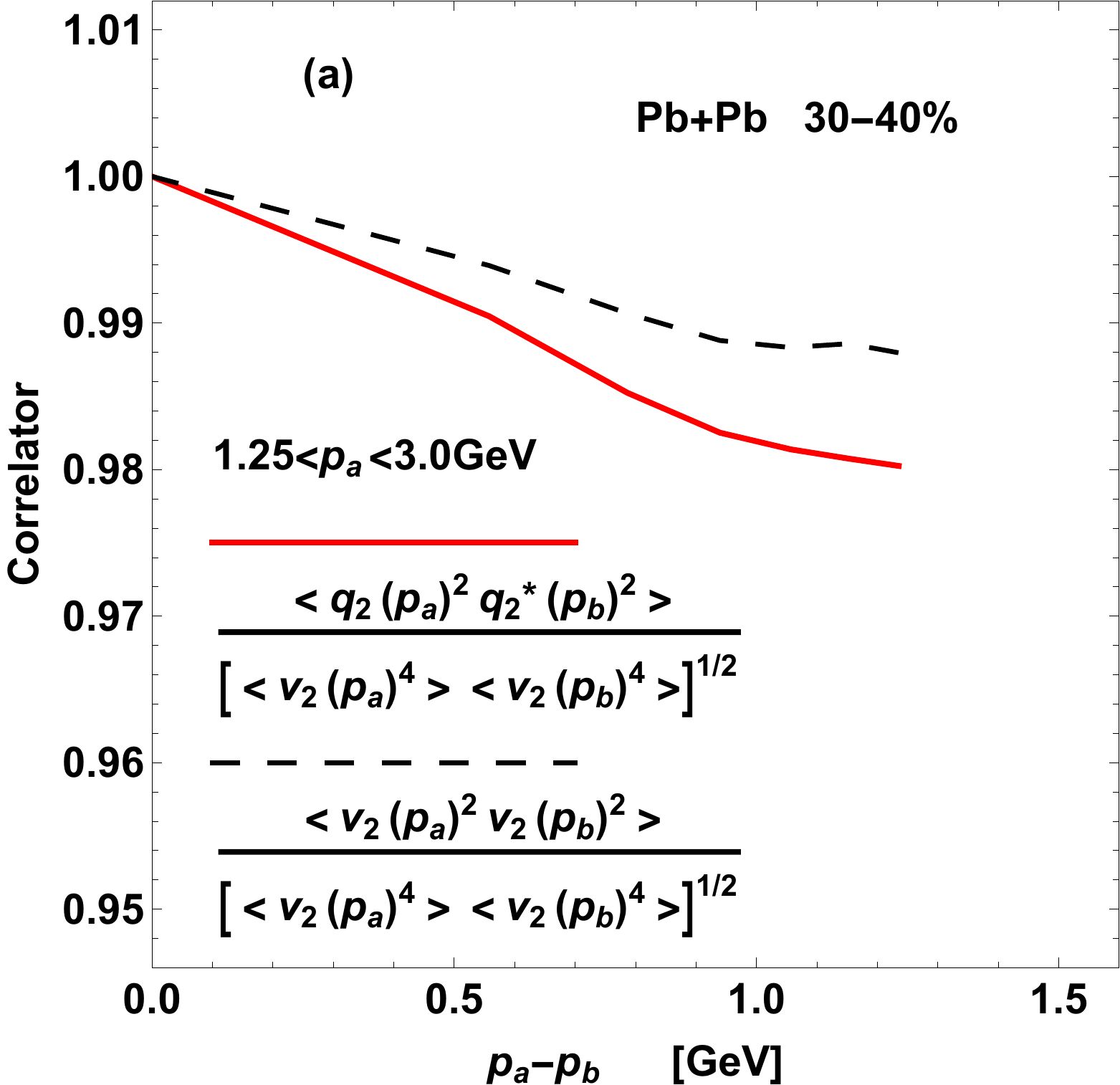}  
\vspace{4mm}

\includegraphics[angle=0,width=0.41 \textwidth]{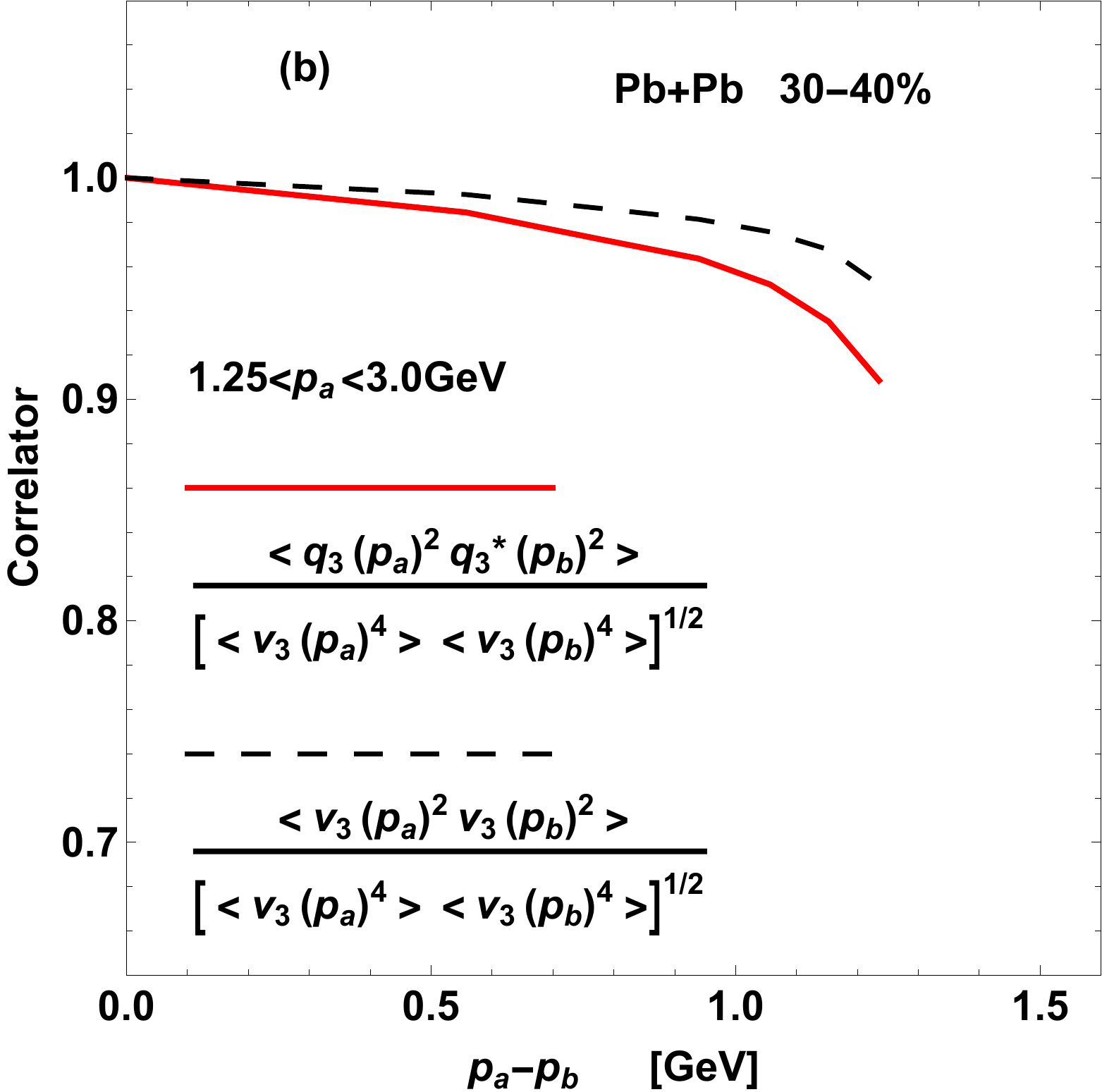}  
\end{center}
\vspace{-5mm}
\caption{Same as Fig. \ref{fig:k205} but for centrality 30-40\%.
\label{fig:k23040}} 
\end{figure}

The angle and magnitude factorization breaking coefficients discussed in Sect. \ref{sec:angmag} 
given by Eqs. \ref{eq:dvnang} and \ref{eq:dvnmag} cannot be measured experimentally.
The angle decorrelation in pseudorapidity  can  been measured separately  using a $4$-bin 
correlator \cite{Aaboud:2017tql,Jia:2017kdq}. That correlator involves $4$ $q_n$ vectors and  should be 
compared to the correlator of the square of the  $q_n$ vectors in two bins.
 It was found that the flow decorrelation (involving flow magnitude and flow angle decorrelation combined) 
is twice as strong
 than the flow angle decorrelation alone, both in experiment \cite{Aaboud:2017tql}
 and in model calculations  \cite{Bozek:2017qir,Wu:2018cpc}.

The situation is slightly different for the 
 decorrelation of the harmonic flow in transverse momentum. One can define a correlator measuring the
 decorrelation of the flow magnitude squared
\begin{equation}
r_{n}^{v_n^2}(p_a,p_b)=\frac{\langle v_n(p_a)^2 v_n(p_b)^2\rangle }{\sqrt{ \langle v_n(p_a)^4 \rangle\langle v_n(p_b)^4 \rangle}} \ .
\label{eq:dvmag2}
\end{equation}
The above correlator can be compared to the correlator of flow vector squared
 $r_{n|n;2}(p_a,p_b)$ (\ref{eq:hiq}).  Both correlators $r_{n}^{v_n^2}(p_a,p_b)$
and $r_{n|n;2}(p_a,p_b)$   involve averages of $4$ $q_n$ vectors and  both correlators can 
be measured in the experiment. The first one measures the flow magnitude decorrelation alone, while the second one measures the flow magnitude and flow angle decorrelations combined.

The predictions for the two correlators for two centralities are presented in Figs. \ref{fig:k205} 
and \ref{fig:k23040}. The decorrelation of the flow magnitude is significantly smaller 
than the flow decorrelation.  Both for the elliptic and triangular flows,
 I find 
 that the flow magnitude decorrelation accounts for roughly half of the total flow decorrelation
\begin{equation}
1-r_{n}^{v_n^2}(p_a,p_b) \simeq \frac{1}{2} \left( 1- r_{n|n;2}(p_a,p_b) \right) \ \ .
\label{eq:simfac}
\end{equation}

\section{Principal component analysis \label{sec:pca}}

\begin{figure}
\begin{center}
\includegraphics[angle=0,width=0.41 \textwidth]{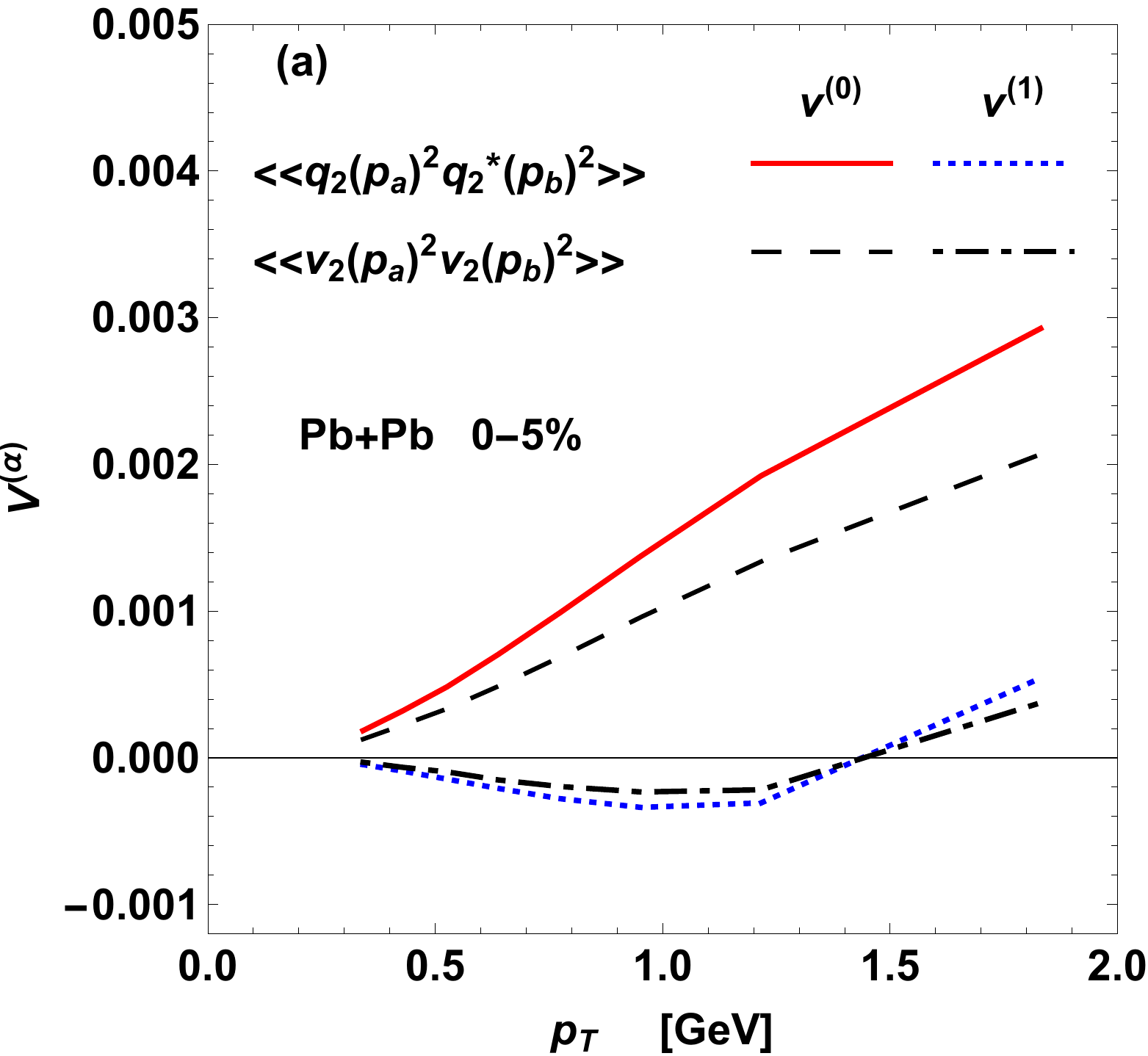}
\vspace{4mm}  

\includegraphics[angle=0,width=0.41 \textwidth]{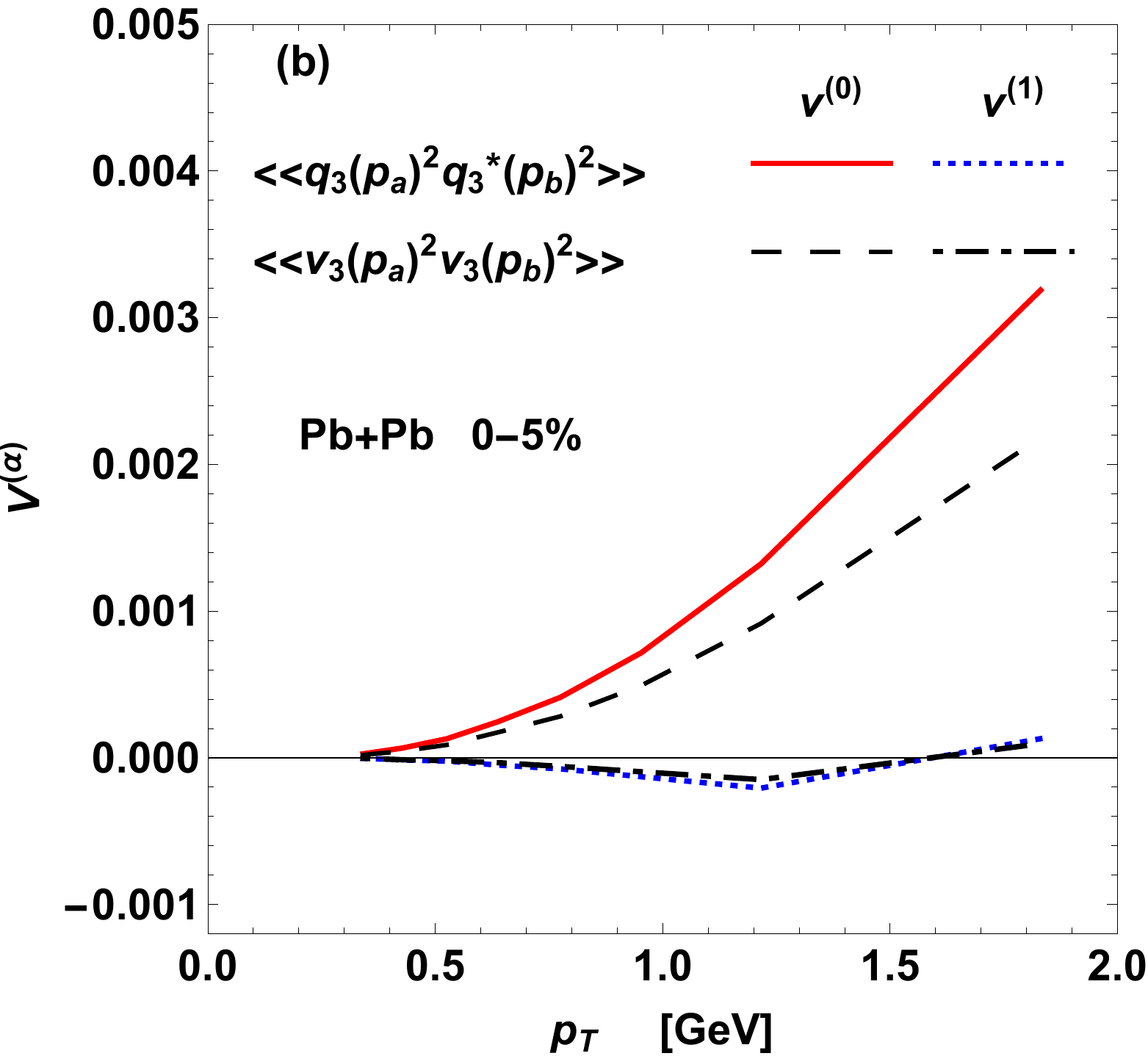}  
\end{center}
\vspace{-5mm}
\caption{Leading (solid line) and subleading (dashed line) eigenmodes of the
correlation matrix of the square of the harmonic flow at two transverse 
momenta.  The dotted and dash-dotted lines denote the leading and subleading 
eigenmodes for the correlation matrix of the square of the magnitude of the harmonic flow. Panel (a) and (b) present results for the elliptic and triangular flow respectively. Pb+Pb collisions at $\sqrt{s_{NN}}=5.02$~TeV for centrality 0-5\%.
\label{fig:vec05}} 
\end{figure}

\begin{figure}
\begin{center}
\includegraphics[angle=0,width=0.41 \textwidth]{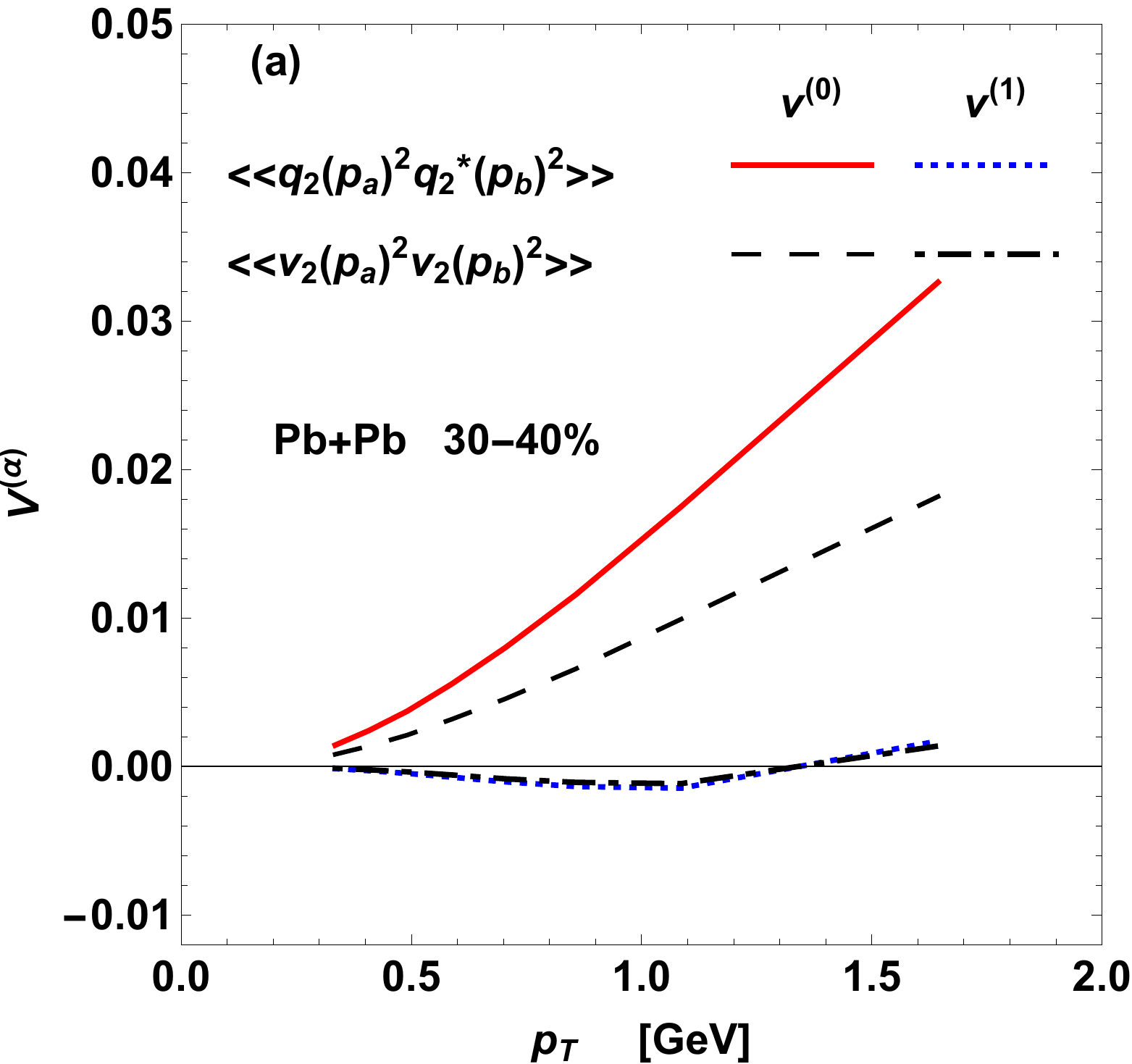}  
\vspace{4mm}

\includegraphics[angle=0,width=0.41 \textwidth]{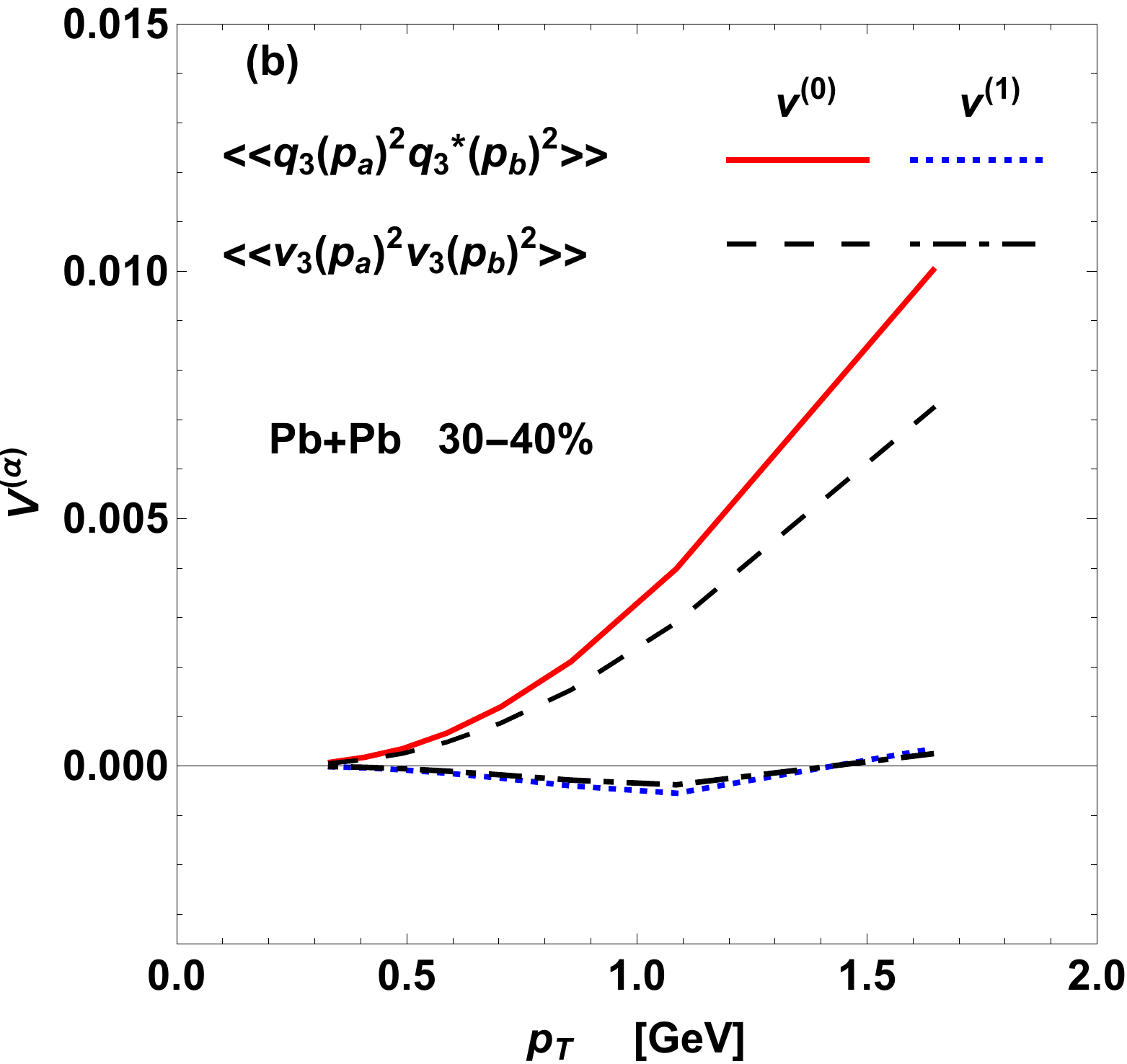}  
\end{center}
\vspace{-5mm}
\caption{Same as Fig. \ref{fig:vec05} but for centrality 30-40\%.
\label{fig:vec3040}} 
\end{figure}

The correlation matrix of the harmonic flow 
can be decomposed into its principal components \cite{Bhalerao:2014mua}
\begin{eqnarray}
&& \langle\langle q_n(p_a) q_n^\star(p_b) \rangle \rangle
 =  \langle q_n(p_a) q_n^\star(p_b) \rangle -\langle q_n(p_a)  \rangle \langle q_n^\star(p_b) \rangle \nonumber \\
 & & =  \langle q_n(p_a) q_n^\star(p_b)  \rangle  = \nonumber \\
&  & v_n^{(0)}(p_a)v_n^{(0)}(p_b)+
v_n^{(1)}(p_a)v_n^{(1)}(p_b)  + \dots \ , \label{eq:corrmat}
\end{eqnarray}
where the leading eigenmode is $v_n^{(0)}(p_a) \simeq \sqrt{\langle v_n(p_a)^2 \rangle } $.
If the subleading modes can be neglected the correlation matrix factorizes
\begin{equation}
\langle\langle q_n(p_a) q_n^\star(p_b) \rangle \rangle \simeq  \sqrt{\langle v_n(p_a)^2 \rangle\langle v_n(p_b)^2 \rangle } \ .
\end{equation}
The factorization breaking coefficient can be
 written as \cite{Bhalerao:2014mua}
\begin{equation}
r_n(p_a,p_b)=1-\frac{1}{2}\left| \frac{v_n^{(1)}(p_a)}{v_n^{(0)}(p_a)}-  \frac{v_n^{(1)}(p_b)}{v_n^{(0)}(p_b)}\right|^2 \ . 
\end{equation}
The principal component decomposition of the flow  correlation matrix
 (\ref{eq:corrmat})
 carries the information about the flow factorization breaking.

The flow magnitude decorrelation discussed in the previous section involves
a correlation of higher powers of the flow vectors.
One can define the decomposition 
\begin{eqnarray}
& & \langle v_n^2(p_a)v_n^2(p_b) \rangle = \nonumber \\
& & v_{v_n^2}^{(0)}(p_a)v_{v_n^2}^{(0)}(p_b)+
v_{v_n^2}^{(1)}(p_a)v_{v_n^2}^{(1)}(p_b)  + \dots \ .
\label{eq:corv2}
\end{eqnarray}
The flow magnitude factorization breaking is
\begin{equation}
r_n^{v_n^2}(p_a,p_b)=1-\frac{1}{2}\left| \frac{v_{v_n^2}^{(1)}(p_a)}{v_{v_n^2}^{(0)}(p_a)}-  \frac{v_{v_n^2}^{(1)}(p_b)}{v_{v_n^2}^{(0)}(p_b)}\right|^2 \ . 
\end{equation}
Please note that the correlator $\langle v_n^2(p_a)v_n^2(p_b) \rangle$ 
is not a correlation matrix.
The proper correlation matrix for $v_n^2$ is
\begin{eqnarray}
& & \langle \langle v_n^2(p_a)v_n^2(p_b) \rangle \rangle = \nonumber \\
 & &   \langle v_n^2(p_a)v_n^2(p_b) \rangle -  \langle v_n^2(p_a) \rangle \langle v_n^2(p_b) \rangle = \nonumber \\
& & \tilde{v}_{v_n^2}^{(0)}(p_a)\tilde{v}_{v_n^2}^{(0)}(p_b)+
\tilde{v}_{v_n^2}^{(1)}(p_a)\tilde{v}_{v_n^2}^{(1)}(p_b)  + \dots \ .
\label{eq:corprop}
\end{eqnarray}
The eigenmode decompositions of the two matrices (\ref{eq:corv2})  and (\ref{eq:corprop}) are related with 
\begin{equation}
\tilde{v}_{v_n^2}^{(1)}(p_a)\simeq {v}_{v_n^2}^{(1)}(p_a) \ \ .
\label{eq:eqsubl}
\end{equation} On the other hand, the dominance of the leading mode gives
\begin{equation}
v_n^{(0)}(p_a)\simeq v_{v_n^2}^{(0)}(p_a) \ .
\label{eq:eqlead}
\end{equation}

In Figs. \ref{fig:vec05} and \ref{fig:vec3040} are shown the eigenmodes for the 
matrices $\langle\langle q_n(p_a) q_n^\star(p_b) \rangle \rangle$
and $\langle \langle  v_n^2(p_a)v_n^2(p_b) \rangle  \rangle$ (due to the relations
(\ref{eq:eqsubl}) and (\ref{eq:eqlead}) the eigenmodes for $\langle   v_n^2(p_a)v_n^2(p_b)   \rangle$ overlap with the curves on the plot).  The subleading mode is much smaller than the leading one, which is consistent with  small
factorization breaking. The subleading modes are similar in shape
\begin{equation}
v_{v_n^2}^{(1}(p_a)\simeq \tilde{v}_{v_n^2}^{(1)}(p_a)\simeq 0.7 v_n^{(1)}(p_a) \  , 
\end{equation}
which leads to  a similar shape of the factorization breaking of the flow and of the flow magnitude (Eq. \ref{eq:simfac}).

\section{Summary}

The factorization breaking of  harmonic flow at two 
different transverse momenta is studied in the hydrodynamic model. 
In the model the decorrelation  of the flow angle
 and of the  flow magnitude  is calculated. The flow angle decorrelation 
is strongly correlated with the overall flow magnitude in an event.
A way to  measure this correlation in experiment is discussed, 
and predictions
are made within the hydrodynamic model.

The separate decorrelation of the flow angle and flow magnitude observed
 in the model cannot
be measured in experiment using two-particle correlation. The flow magnitude 
decorrelation could be measured in experiment using a $4$-particle correlations.
The hydrodynamic model with Monte Carlo Glauber initial conditions 
predicts  that the
 flow magnitude decorrelation is about one half of the 
overall flow decorrelation. The difference in the flow
factorization breaking  or  the flow magnitude factorization breaking
can be studied using the principal component analysis of the relevant 
$4$-particle
correlation matrices. The hierarchy of the eigenmodes in the 
principal component analysis is consistent with results on 
 factorization breaking.

\section*{Acknowledgments}

Research  supported by the  AGH UST statutory funds, by the National
Science Centre grant 2015/17/B/ST2/00101, as well as by PL-Grid Infrastructure. 

\bibliography{../hydr}

\begin{thebibliography}{10}%
\makeatletter
\providecommand \@ifxundefined [1]{%
 \ifx #1\undefined \expandafter \@firstoftwo
 \else \expandafter \@secondoftwo
\fi
}%
\providecommand \@ifnum [1]{%
 \ifnum #1\expandafter \@firstoftwo
 \else \expandafter \@secondoftwo
\fi
}%
\providecommand \enquote [1]{``#1''}%
\providecommand \bibnamefont  [1]{#1}%
\providecommand \bibfnamefont [1]{#1}%
\providecommand \citenamefont [1]{#1}%
\providecommand\href[0]{\@sanitize\@href}%
\providecommand\@href[1]{\endgroup\@@startlink{#1}\endgroup\@@href}%
\providecommand\@@href[1]{#1\@@endlink}%
\providecommand \@sanitize [0]{\begingroup\catcode`\&12\catcode`\#12\relax}%
\@ifxundefined \pdfoutput {\@firstoftwo}{%
 \@ifnum{\z@=\pdfoutput}{\@firstoftwo}{\@secondoftwo}%
}{%
 \providecommand\@@startlink[1]{\leavevmode\special{html:<a href="#1">}}%
 \providecommand\@@endlink[0]{\special{html:</a>}}%
}{%
 \providecommand\@@startlink[1]{%
  \leavevmode
  \pdfstartlink
   attr{/Border[0 0 1 ]/H/I/C[0 1 1]}%
   user{/Subtype/Link/A<</Type/Action/S/URI/URI(#1)>>}%
  \relax
 }%
 \providecommand\@@endlink[0]{\pdfendlink}%
}%
\providecommand \url  [0]{\begingroup\@sanitize \@url }%
\providecommand \@url [1]{\endgroup\@href {#1}{\urlprefix}}%
\providecommand \urlprefix [0]{URL }%
\providecommand \Eprint[0]{\href }%
\@ifxundefined \urlstyle {%
  \providecommand \doi [1]{doi:\discretionary{}{}{}#1}%
}{%
  \providecommand \doi [0]{doi:\discretionary{}{}{}\begingroup
  \urlstyle{rm}\Url }%
}%
\providecommand \doibase [0]{http://dx.doi.org/}%
\providecommand \Doi[1]{\href{\doibase#1}}%
\providecommand \bibAnnote [3]{%
  \BibitemShut{#1}%
  \begin{quotation}\noindent
    \textsc{Key:}\ #2\\\textsc{Annotation:}\ #3%
  \end{quotation}%
}%
\providecommand \bibAnnoteFile [2]{%
  \IfFileExists{#2}{\bibAnnote {#1} {#2} {\input{#2}}}{}%
}%
\providecommand \typeout [0]{\immediate \write \m@ne }%
\providecommand \selectlanguage [0]{\@gobble}%
\providecommand \bibinfo [0]{\@secondoftwo}%
\providecommand \bibfield [0]{\@secondoftwo}%
\providecommand \translation [1]{[#1]}%
\providecommand \BibitemOpen[0]{}%
\providecommand \bibitemStop [0]{}%
\providecommand \bibitemNoStop [0]{.\EOS\space}%
\providecommand \EOS [0]{\spacefactor3000\relax}%
\providecommand \BibitemShut [1]{\csname bibitem#1\endcsname}%
\bibitem{Heinz:2013th}%
  \BibitemOpen
  \bibfield{author}{%
  \bibinfo {author} {\bibfnamefont{U.}~\bibnamefont{Heinz}}\ and\ \bibinfo
  {author} {\bibfnamefont{R.}~\bibnamefont{Snellings}},\ }%
  \bibfield{journal}{%
  \Doi{10.1146/annurev-nucl-102212-170540}{\bibinfo {journal}
  {Ann.Rev.Nucl.Part.Sci.}}\ }%
  \textbf{\bibinfo {volume} {63}},\ \bibinfo {pages} {123} (\bibinfo {year}
  {2013})%
  \bibAnnoteFile{NoStop}{Heinz:2013th}%
\bibitem{Gale:2013da}%
  \BibitemOpen
  \bibfield{author}{%
  \bibinfo {author} {\bibfnamefont{C.}~\bibnamefont{Gale}}, \bibinfo {author}
  {\bibfnamefont{S.}~\bibnamefont{Jeon}},\ and\ \bibinfo {author}
  {\bibfnamefont{B.}~\bibnamefont{Schenke}},\ }%
  \bibfield{journal}{%
  \Doi{10.1142/S0217751X13400113}{\bibinfo {journal} {Int.J.Mod.Phys.}}\ }%
  \textbf{\bibinfo {volume} {A28}},\ \bibinfo {pages} {1340011} (\bibinfo
  {year} {2013})%
  \bibAnnoteFile{NoStop}{Gale:2013da}%
\bibitem{Ollitrault:2010tn}%
  \BibitemOpen
  \bibfield{author}{%
  \bibinfo {author} {\bibfnamefont{J.-Y.}\ \bibnamefont{Ollitrault}},\ }%
  \bibfield{journal}{%
  \Doi{10.1088/1742-6596/312/1/012002}{\bibinfo {journal} {J. Phys. Conf.
  Ser.}}\ }%
  \textbf{\bibinfo {volume} {312}},\ \bibinfo {pages} {012002} (\bibinfo {year}
  {2011})%
  \bibAnnoteFile{NoStop}{Ollitrault:2010tn}%
\bibitem{Bozek:2010vz}%
  \BibitemOpen
  \bibfield{author}{%
  \bibinfo {author} {\bibfnamefont{P.}~\bibnamefont{Bo\.zek}}, \bibinfo
  {author} {\bibfnamefont{W.}~\bibnamefont{Broniowski}},\ and\ \bibinfo
  {author} {\bibfnamefont{J.}~\bibnamefont{Moreira}},\ }%
  \bibfield{journal}{%
  \Doi{10.1103/PhysRevC.83.034911}{\bibinfo {journal} {Phys. Rev.}}\ }%
  \textbf{\bibinfo {volume} {C83}},\ \bibinfo {pages} {034911} (\bibinfo {year}
  {2011})%
  \bibAnnoteFile{NoStop}{Bozek:2010vz}%
\bibitem{Gardim:2012im}%
  \BibitemOpen
  \bibfield{author}{%
  \bibinfo {author} {\bibfnamefont{F.~G.}\ \bibnamefont{Gardim}}, \bibinfo
  {author} {\bibfnamefont{F.}~\bibnamefont{Grassi}}, \bibinfo {author}
  {\bibfnamefont{M.}~\bibnamefont{Luzum}},\ and\ \bibinfo {author}
  {\bibfnamefont{J.-Y.}\ \bibnamefont{Ollitrault}},\ }%
  \bibfield{journal}{%
  \Doi{10.1103/PhysRevC.87.031901}{\bibinfo {journal} {Phys.Rev.}}\ }%
  \textbf{\bibinfo {volume} {C87}},\ \bibinfo {pages} {031901} (\bibinfo {year}
  {2013})%
  \bibAnnoteFile{NoStop}{Gardim:2012im}%
\bibitem{Acharya:2017ino}%
  \BibitemOpen
  \bibfield{author}{%
  \bibinfo {author} {\bibfnamefont{S.}~\bibnamefont{Acharya}} \emph{et~al.}
  (\bibinfo {collaboration} {ALICE Collaboration}),\ }%
  \bibfield{journal}{%
  \Doi{10.1007/JHEP09(2017)032}{\bibinfo {journal} {JHEP}}\ }%
  \textbf{\bibinfo {volume} {09}},\ \bibinfo {pages} {032} (\bibinfo {year}
  {2017})%
  \bibAnnoteFile{NoStop}{Acharya:2017ino}%
\bibitem{Aad:2014lta}%
  \BibitemOpen
  \bibfield{author}{%
  \bibinfo {author} {\bibfnamefont{G.}~\bibnamefont{Aad}} \emph{et~al.}
  (\bibinfo {collaboration} {ATLAS Collaboration}),\ }%
  \bibfield{journal}{%
  \Doi{10.1103/PhysRevC.90.044906}{\bibinfo {journal} {Phys.Rev.}}\ }%
  \textbf{\bibinfo {volume} {C90}},\ \bibinfo {pages} {044906} (\bibinfo {year}
  {2014})%
  \bibAnnoteFile{NoStop}{Aad:2014lta}%
\bibitem{Khachatryan:2015oea}%
  \BibitemOpen
  \bibfield{author}{%
  \bibinfo {author} {\bibfnamefont{V.}~\bibnamefont{Khachatryan}} \emph{et~al.}
  (\bibinfo {collaboration} {CMS Collaboration}),\ }%
  \bibfield{journal}{%
  \Doi{10.1103/PhysRevC.92.034911}{\bibinfo {journal} {Phys. Rev.}}\ }%
  \textbf{\bibinfo {volume} {C92}},\ \bibinfo {pages} {034911} (\bibinfo {year}
  {2015})%
  \bibAnnoteFile{NoStop}{Khachatryan:2015oea}%
\bibitem{Aaboud:2017tql}%
  \BibitemOpen
  \bibfield{author}{%
  \bibinfo {author} {\bibfnamefont{M.}~\bibnamefont{Aaboud}} \emph{et~al.}
  (\bibinfo {collaboration} {ATLAS}),\ }%
  \bibfield{journal}{%
  \Doi{10.1140/epjc/s10052-018-5605-7}{\bibinfo {journal} {Eur. Phys. J.}}\ }%
  \textbf{\bibinfo {volume} {C78}},\ \bibinfo {pages} {142} (\bibinfo {year}
  {2018})%
  \bibAnnoteFile{NoStop}{Aaboud:2017tql}%
\bibitem{Heinz:2013bua}%
  \BibitemOpen
  \bibfield{author}{%
  \bibinfo {author} {\bibfnamefont{U.}~\bibnamefont{Heinz}}, \bibinfo {author}
  {\bibfnamefont{Z.}~\bibnamefont{Qiu}},\ and\ \bibinfo {author}
  {\bibfnamefont{C.}~\bibnamefont{Shen}},\ }%
  \bibfield{journal}{%
  \Doi{10.1103/PhysRevC.87.034913}{\bibinfo {journal} {Phys.Rev.}}\ }%
  \textbf{\bibinfo {volume} {C87}},\ \bibinfo {pages} {034913} (\bibinfo {year}
  {2013})%
  \bibAnnoteFile{NoStop}{Heinz:2013bua}%
\bibitem{Kozlov:2014fqa}%
  \BibitemOpen
  \bibfield{author}{%
  \bibinfo {author} {\bibfnamefont{I.}~\bibnamefont{Kozlov}}, \bibinfo {author}
  {\bibfnamefont{M.}~\bibnamefont{Luzum}}, \bibinfo {author}
  {\bibfnamefont{G.}~\bibnamefont{Denicol}}, \bibinfo {author}
  {\bibfnamefont{S.}~\bibnamefont{Jeon}},\ and\ \bibinfo {author}
  {\bibfnamefont{C.}~\bibnamefont{Gale}}}%
   (\bibinfo {year} {2014}),\
  \Eprint{http://arxiv.org/abs/1405.3976}{arXiv:1405.3976 [nucl-th]}%
  \bibAnnoteFile{NoStop}{Kozlov:2014fqa}%
\bibitem{Lin:2004en}%
  \BibitemOpen
  \bibfield{author}{%
  \bibinfo {author} {\bibfnamefont{Z.-W.}\ \bibnamefont{Lin}}, \bibinfo
  {author} {\bibfnamefont{C.~M.}\ \bibnamefont{Ko}}, \bibinfo {author}
  {\bibfnamefont{B.-A.}\ \bibnamefont{Li}}, \bibinfo {author}
  {\bibfnamefont{B.}~\bibnamefont{Zhang}},\ and\ \bibinfo {author}
  {\bibfnamefont{S.}~\bibnamefont{Pal}},\ }%
  \bibfield{journal}{%
  \Doi{10.1103/PhysRevC.72.064901}{\bibinfo {journal} {Phys. Rev.}}\ }%
  \textbf{\bibinfo {volume} {C72}},\ \bibinfo {pages} {064901} (\bibinfo {year}
  {2005})%
  \bibAnnoteFile{NoStop}{Lin:2004en}%
\bibitem{Pang:2015zrq}%
  \BibitemOpen
  \bibfield{author}{%
  \bibinfo {author} {\bibfnamefont{L.-G.}\ \bibnamefont{Pang}}, \bibinfo
  {author} {\bibfnamefont{H.}~\bibnamefont{Petersen}}, \bibinfo {author}
  {\bibfnamefont{G.-Y.}\ \bibnamefont{Qin}}, \bibinfo {author}
  {\bibfnamefont{V.}~\bibnamefont{Roy}},\ and\ \bibinfo {author}
  {\bibfnamefont{X.-N.}\ \bibnamefont{Wang}},\ }%
  \bibfield{journal}{%
  \Doi{10.1140/epja/i2016-16097-x}{\bibinfo {journal} {Eur. Phys. J.}}\ }%
  \textbf{\bibinfo {volume} {A52}},\ \bibinfo {pages} {97} (\bibinfo {year}
  {2016})%
  \bibAnnoteFile{NoStop}{Pang:2015zrq}%
\bibitem{Xiao:2015dma}%
  \BibitemOpen
  \bibfield{author}{%
  \bibinfo {author} {\bibfnamefont{K.}~\bibnamefont{Xiao}}, \bibinfo {author}
  {\bibfnamefont{L.}~\bibnamefont{Yi}}, \bibinfo {author}
  {\bibfnamefont{F.}~\bibnamefont{Liu}},\ and\ \bibinfo {author}
  {\bibfnamefont{F.}~\bibnamefont{Wang}},\ }%
  \bibfield{journal}{%
  \Doi{10.1103/PhysRevC.94.024905}{\bibinfo {journal} {Phys. Rev.}}\ }%
  \textbf{\bibinfo {volume} {C94}},\ \bibinfo {pages} {024905} (\bibinfo {year}
  {2016})%
  \bibAnnoteFile{NoStop}{Xiao:2015dma}%
\bibitem{Jia:2017kdq}%
  \BibitemOpen
  \bibfield{author}{%
  \bibinfo {author} {\bibfnamefont{J.}~\bibnamefont{Jia}}, \bibinfo {author}
  {\bibfnamefont{P.}~\bibnamefont{Huo}}, \bibinfo {author}
  {\bibfnamefont{G.}~\bibnamefont{Ma}},\ and\ \bibinfo {author}
  {\bibfnamefont{M.}~\bibnamefont{Nie}},\ }%
  \bibfield{journal}{%
  \Doi{10.1088/1361-6471/aa74c3}{\bibinfo {journal} {J. Phys.}}\ }%
  \textbf{\bibinfo {volume} {G44}},\ \bibinfo {pages} {075106} (\bibinfo {year}
  {2017})%
  \bibAnnoteFile{NoStop}{Jia:2017kdq}%
\bibitem{Bozek:2017qir}%
  \BibitemOpen
  \bibfield{author}{%
  \bibinfo {author} {\bibfnamefont{P.}~\bibnamefont{Bo{\.z}ek}}\ and\ \bibinfo
  {author} {\bibfnamefont{W.}~\bibnamefont{Broniowski}},\ }%
  \bibfield{journal}{%
  \Doi{10.1103/PhysRevC.97.034913}{\bibinfo {journal} {Phys. Rev.}}\ }%
  \textbf{\bibinfo {volume} {C97}},\ \bibinfo {pages} {034913} (\bibinfo {year}
  {2018})%
  \bibAnnoteFile{NoStop}{Bozek:2017qir}%
\bibitem{Wu:2018cpc}%
  \BibitemOpen
  \bibfield{author}{%
  \bibinfo {author} {\bibfnamefont{X.-Y.}\ \bibnamefont{Wu}}, \bibinfo {author}
  {\bibfnamefont{L.-G.}\ \bibnamefont{Pang}}, \bibinfo {author}
  {\bibfnamefont{G.-Y.}\ \bibnamefont{Qin}},\ and\ \bibinfo {author}
  {\bibfnamefont{X.-N.}\ \bibnamefont{Wang}}}%
   (\bibinfo {year} {2018}),\
  \Eprint{http://arxiv.org/abs/1805.03762}{arXiv:1805.03762 [nucl-th]}%
  \bibAnnoteFile{NoStop}{Wu:2018cpc}%
\bibitem{Schenke:2010rr}%
  \BibitemOpen
  \bibfield{author}{%
  \bibinfo {author} {\bibfnamefont{B.}~\bibnamefont{Schenke}}, \bibinfo
  {author} {\bibfnamefont{S.}~\bibnamefont{Jeon}},\ and\ \bibinfo {author}
  {\bibfnamefont{C.}~\bibnamefont{Gale}},\ }%
  \bibfield{journal}{%
  \Doi{10.1103/PhysRevLett.106.042301}{\bibinfo {journal} {Phys. Rev. Lett.}}\
  }%
  \textbf{\bibinfo {volume} {106}},\ \bibinfo {pages} {042301} (\bibinfo {year}
  {2011})%
  \bibAnnoteFile{NoStop}{Schenke:2010rr}%
\bibitem{Bozek:2011ua}%
  \BibitemOpen
  \bibfield{author}{%
  \bibinfo {author} {\bibfnamefont{P.}~\bibnamefont{Bo\.zek}},\ }%
  \bibfield{journal}{%
  \Doi{10.1103/PhysRevC.85.034901}{\bibinfo {journal} {Phys. Rev.}}\ }%
  \textbf{\bibinfo {volume} {C85}},\ \bibinfo {pages} {034901} (\bibinfo {year}
  {2012})%
  \bibAnnoteFile{NoStop}{Bozek:2011ua}%
\bibitem{Chojnacki:2011hb}%
  \BibitemOpen
  \bibfield{author}{%
  \bibinfo {author} {\bibfnamefont{M.}~\bibnamefont{Chojnacki}}, \bibinfo
  {author} {\bibfnamefont{A.}~\bibnamefont{Kisiel}}, \bibinfo {author}
  {\bibfnamefont{W.}~\bibnamefont{Florkowski}},\ and\ \bibinfo {author}
  {\bibfnamefont{W.}~\bibnamefont{Broniowski}},\ }%
  \bibfield{journal}{%
  \Doi{10.1016/j.cpc.2011.11.018}{\bibinfo {journal} {Comput. Phys. Commun.}}\
  }%
  \textbf{\bibinfo {volume} {183}},\ \bibinfo {pages} {746} (\bibinfo {year}
  {2012})%
  \bibAnnoteFile{NoStop}{Chojnacki:2011hb}%
\bibitem{Bozek:2017thv}%
  \BibitemOpen
  \bibfield{author}{%
  \bibinfo {author} {\bibfnamefont{P.}~\bibnamefont{Bo{\.z}ek}},\ }%
  \bibfield{journal}{%
  \Doi{10.1103/PhysRevC.97.034905}{\bibinfo {journal} {Phys. Rev.}}\ }%
  \textbf{\bibinfo {volume} {C97}},\ \bibinfo {pages} {034905} (\bibinfo {year}
  {2018})%
  \bibAnnoteFile{NoStop}{Bozek:2017thv}%
\bibitem{Jia:2014vja}%
  \BibitemOpen
  \bibfield{author}{%
  \bibinfo {author} {\bibfnamefont{J.}~\bibnamefont{Jia}}\ and\ \bibinfo
  {author} {\bibfnamefont{P.}~\bibnamefont{Huo}},\ }%
  \bibfield{journal}{%
  \Doi{10.1103/PhysRevC.90.034905}{\bibinfo {journal} {Phys.Rev.}}\ }%
  \textbf{\bibinfo {volume} {C90}},\ \bibinfo {pages} {034905} (\bibinfo {year}
  {2014})%
  \bibAnnoteFile{NoStop}{Jia:2014vja}%
\bibitem{Bhalerao:2014mua}%
  \BibitemOpen
  \bibfield{author}{%
  \bibinfo {author} {\bibfnamefont{R.~S.}\ \bibnamefont{Bhalerao}}, \bibinfo
  {author} {\bibfnamefont{J.-Y.}\ \bibnamefont{Ollitrault}}, \bibinfo {author}
  {\bibfnamefont{S.}~\bibnamefont{Pal}},\ and\ \bibinfo {author}
  {\bibfnamefont{D.}~\bibnamefont{Teaney}},\ }%
  \bibfield{journal}{%
  \Doi{10.1103/PhysRevLett.114.152301}{\bibinfo {journal} {Phys. Rev. Lett.}}\
  }%
  \textbf{\bibinfo {volume} {114}},\ \bibinfo {pages} {152301} (\bibinfo {year}
  {2015})%
  \bibAnnoteFile{NoStop}{Bhalerao:2014mua}%
\end{thebibliography}%

\end{document}